\documentclass[a4paper,12pt]{article}

\usepackage{amsmath}
\usepackage{graphicx}
\usepackage{setspace} 
\usepackage{natbib}
\bibpunct{(}{)}{;}{author-year}{}{,}
\usepackage{genetics}
\usepackage{fullpage}
\usepackage{url}
\usepackage[font={footnotesize,sf},labelfont=bf,labelsep=space,justification=justified]{caption}

\date{}

\newcommand\ind{\bot\hspace*{-6pt}\bot}
\newcommand\jind[2]{#1\ind#2}

\title{Mapping eQTL networks with \\ mixed graphical Markov models}
\author{Inma Tur\thanks{Dept. of Experimental and Health Sciences, Universitat Pompeu Fabra, E-08003 Barcelona, Spain} \thanks{Research Programme on Biomedical Informatics, Institut Hospital del Mar d'Investigacions M\`ediques, E-08003 Barcelona, Spain} \and Alberto Roverato\thanks{Dept. of Statistical Sciences, Universit\`a di Bologna, I-40126 Bologna, Italy} \and Robert Castelo\footnotemark[1] \footnotemark[2]}

\begin{document}

\maketitle

\clearpage

\ \\

\vspace{1cm}
\noindent
Short running title: Mapping eQTL networks with mixed GMMs


\bigskip
\noindent
\begin{tabular}{ll}
\hspace{-0.27cm}Corresponding author: & Robert Castelo \\
                      & GRIB-UPF-PRBB \\
                      & Dr. Aiguader 88 \\
                      & E-08003 Barcelona \\
                      & Spain \\
                      & Telf.: {\tt + 34 933 160 514} \\
                      & Email: {\tt robert.castelo@upf.edu}
\end{tabular}

\clearpage

\begin{abstract}
Expression quantitative trait loci (eQTL) mapping constitutes a challenging
problem due to, among other reasons, the high-dimensional multivariate nature
of gene-expression traits. Next to the expression heterogeneity produced by
confounding factors and other sources of unwanted variation, indirect effects
spread throughout genes as a result of genetic, molecular and environmental
perturbations. From a multivariate perspective one would like to adjust for
the effect of all of these factors to end up with a network of direct
associations connecting the path from genotype to phenotype. In this paper we
approach this challenge with mixed graphical Markov models, higher-order
conditional independences and $q$-order correlation graphs. These models show
that additive genetic effects propagate through the network as function of
gene-gene correlations. Our estimation of the eQTL network underlying a
well-studied yeast data set leads to a sparse structure with more direct genetic
and regulatory associations that enable a straightforward comparison of the
genetic control of gene expression across chromosomes. Interestingly, it also
reveals that eQTLs explain most of the expression variability of network hub
genes.
\end{abstract}

\clearpage

\section*{Introduction}

The simultaneous assay of gene-expression profiling and genotyping on the
same samples with high-throughput technologies provides one of the
primary types of integrative genomics data sets, the so-called {\it genetical
genomics} data \citep{JansenNap:2001}. First studies producing such data showed
that, for many genes, gene expression is an heritable trait
\citep{BremKruglyak:2002}. Following this observation, it soon became evident
that gene expression may act as an intermediate data tier that can potentially
increase our power to map the genetic component of phenotypic variability in
complex traits, such as human disease \citep{SchadtFriend:2003}. The genetic
variants associated with each of these thousands of molecular phenotypes are known
as expression quantitative trait loci (eQTL) and they can be broadly categorized
into {\it cis}-acting and {\it trans}-acting associations, depending on their
location relative to the gene whose expression levels map to the
eQTL\footnote{We use here the terms {\it cis} and {\it trans} to refer to what
is also known as {\it local} and {\it distant} QTLs \citep{RockmanKruglyak:2006},
respectively.}. Because the relative concentration of RNA molecules reflects
functional relationships between genes, overlaying the correlation structure
of gene expression on the eQTL associations provides an {\it eQTL network} which
can help to approach the problem of reverse engineering the genotype-phenotype
map with natural variation \citep{Rockman:2008}.

A straightforward way to map eQTL networks to the genome is by treating gene
expression profiles as independent continuous traits and applying classical
QTL mapping techniques such as single marker regression \citep{TessonJansen:2009}.
However, differently than higher-level phenotypes such as disease onset, adult
height or yeast growth rates, gene expression is a high-dimensional multivariate
trait involving measurements from thousands of genes coordinately acting under
complex molecular regulatory programs. This feature makes eQTL mapping
a challenging problem for, at least, two reasons. One is that gene expression
profiles can be highly correlated as a product of gene regulation, thereby
complicating the distinction between direct and indirect effects when marginally
inspecting eQTL associations that only involve one gene at a time. The other is
that high-throughput expression profiling can be very sensitive to nonbiological
factors of variation such as batch effects \citep{LeekStorey:2007,LeekIrizarry:2010},
introducing heterogeneity and spurious correlations between gene expression
measurements. These artifacts may compromise the statistical power to map
truly biological eQTLs \citep{StegleWinn:2010} or show up as interesting genetic
switches with broad pleiotropic effects affecting a large number of genes,
commonly known as eQTL hotspots \citep{LeekStorey:2007, BreitlingJansen:2008}.
These problems can be addressed by estimating and including the confounding
factors in the model as main \citep{LeekStorey:2007,StegleWinn:2010} or mixed
effects \citep{Kang:2008,Listgarten:2010} and restricting the eQTL search to
{\it cis}-acting variants located in the regulatory regions of the gene to which they
are associated \citep{MontgomeryDermitzakis:2010}.

Yet, {\it trans}-acting eQTL have proven to be crucial to our understanding of
complex regulatory mechanisms. A canonical example is locus control regions
\citep{LiStam:2002} that enhance the expression of distal genes under
tissue-specific conditions such as the one affecting the human $\beta$-globin
locus. Recent contributions have shown that {\it trans}-acting mechanisms often
mediate the genetic basis of disease \citep{WestraFranke:2013}.

The importance of identifying nonspurious {\it trans}-acting eQTLs has been
widely recognized and a large number of approaches that aim to address the
problems described above exist in the literature. They can be broadly
categorized into those extending univariate single marker regression models to
adjust for confounding effects \citep{LeekStorey:2007,StegleWinn:2010,Listgarten:2010}
and those using multivariate approaches. The latter can be further categorized into
Bayesian networks using conditional mutual information with constraint-based algorithms \citep{Zhu:2004},
empirical Bayes hierarchical mixtures \citep{KendziorskiAttie:2006},
directed versions of the PC algorithm \citep{Neto:2008},
structural equation models \citep{Liu:2008},
sparse partial least squares \citep{ChunKeles:2009},
fused lasso regression methods \citep{Kim:2009},
random forests \citep{MichaelsonBeyer:2010},
mixed Bayesian networks using the Bayesian information criterion (BIC) with homogeneous conditional Gaussian regression models \citep{Neto:2010},
mixed graphical Markov models restricted to tree network topologies \citep{Edwards:2010},
sparse factor analysis \citep{Parts:2011},
and
conditional independence tests of order one \citep{Bing:2005,Chen:2007,Kang:2010,Neto:2013}.

Mixed graphical Markov models and conditional independence constitute a natural
extension of classical QTL mapping to multivariate phenotype vectors. This enables
a smooth transition from mapping single phenotypes to building eQTL networks
providing an easier statistical interpretation of the resulting associations.
However, a main shortcoming of currently available methods based on mixed graphical
Markov models and conditional independence is that they are restricted to conditioning
on one other gene to disentangle direct and indirect relationships.

Using higher-order conditional independences on high-dimensional data, such as
the one produced by genetical genomics experiments, is not trivial. The main
purpose of this article is twofold: first, to show a way to use higher-order
conditional independence and mixed graphical Markov models in eQTL mapping by
means of $q$-order correlation graphs, and second, to demonstrate that this
approach can help to obtain valuable insight into the genetic regulatory
architecture of gene expression in yeast.

\section{Materials and Methods}
\label{sec:methods}

\noindent\newline\textbf{Expression measurements and genotyping}
\label{sec:yeast}

Throughout this article we use a well-studied genetical genomics data set generated
from two yeast strains, a wild-type (RM11-1a) and a lab strain (BY4716) that were
crossed to generate 112 segregants, which were genotyped and whose gene expression
was profiled using two-channel microarray chips, including a dye-swap \citep{Brem:2005}.
We applied background correction, discarded control probes and normalized the
raw expression data within and between arrays using the \texttt{limma}
package \citep{Smyth:2003,Ritchie:2007}. To correct for possible dye effects, we
averaged the normalized expression values of the dye-swapped arrays. This first
set of normalized data consisted of 6,216 genes and 2,906 genotype markers, i.e.,
a total of $p=9,122$ features, by $n=112$ samples.

Using the \texttt{geno.table()} and \texttt{findDupMarkers()} functions from the
\texttt{R/qtl} package \citep{Broman:2003} we identified 274 markers with $> 5$\%
missing genotypes and 749 markers with duplicated genotypes in other
260 markers. These 274+749=1,023 markers were discarded from further analysis.
Among the remaining 1,883 we discarded 26 showing Mendelian segregation
distortion at Holm's FWER $< 0.01$. We also removed 75 genes that could not be
found in the March 2014 version of the \texttt{sgdGene} annotation table for the
sacCer3 version of the yeast genome at the UCSC Genome Browser
(\url{http://www.genome.ucsc.edu}). These filtering steps resulted in a final
data set of 1,857 markers and 6,141 genes, i.e., a total of $p=7,998$ features,
by $n=112$ samples. Missing genotypes were treated by complete-case analysis.
The use of marginal distributions in the next section enables this approach.

\noindent\newline\textbf{Software and reproducibility}

The algorithms described in this article are implemented in the open source
R/Bioconductor package \texttt{qpgraph} available for download at
\url{http://www.bioconductor.org}. Scripts with the R code reproducing the
results of this article are available at
\url{http://functionalgenomics.upf.edu/supplements/eQTLmixedGMMs}.

\noindent\newline\textbf{Mixed Graphical Markov models of eQTL networks}
\label{sec:hmgms}

We assumed that gene expression forms a $p$-multivariate
sample following a conditional Gaussian distribution given the joint probability
of all genetic variants. Under this assumption a sensible model for an eQTL
network is a mixed graphical Markov model -GMM- \citep{Lauritzen:1989}. A
mixed GMM enables the integration of the joint distribution of discrete genotypes
with the joint distribution of continuous expression measurements in a single
multivariate statistical model satisfying a set of restrictions of conditional
independence encoded by means of a graph; see \citep{Lauritzen:1996,Edwards:2000}
for a comprehensive description of this type of statistical model.

Here, we review part of the mixed GMM theory required for
this paper. Mixed GMMs are statistical models representing probability
distributions involving discrete random variables (r.v.'s), denoted by
$I_\delta$ with $\delta \in \Delta$, and continuous r.v.'s, denoted by
$Y_\gamma$ with $\gamma \in \Gamma$. This class of GMMs are determined by
marked graphs $G=(V,E)$ with $p$ marked vertices $V=\Delta \cup \Gamma$,
and edge set $E\subseteq V\times V$. Vertices $\delta\in\Delta$ are
depicted by solid circles, $\gamma\in\Gamma$ by open ones and the entire
set of them, $V$, index the vector of r.v.'s $X=(I, Y)$. In our
context, continuous r.v.'s $Y$ correspond to genes and discrete r.v.'s $I$
to markers or eQTLs; see Figure~\ref{additiveEffects}A for a graphical
representation of one such mixed GMM. We denote the joint sample space of
$X$ by
\begin{equation}
x = (i, y) = \left\{(i_\delta)_{\delta \in \Delta}, (y_\gamma)_{\gamma \in \Gamma} \right\}\,,
\end{equation}
where $i_\delta$ are discrete values corresponding to genotype alleles from the
marker or eQTL $I_\delta$ and $y_\gamma$ are continuous values of expression
from gene $Y_\gamma$. The set of all possible joint discrete levels $i$ is
denoted as $\mathcal{I}$. Following \citep{Lauritzen:1989}, we assume that the
joint distribution of the variables $X$ is conditional Gaussian (also known as
CG-distribution) with density function:
\begin{equation}
f(x) = f(i,y) = p(i) |2\pi \Sigma(i)|^{-\frac{1}{2}}\times \exp\left\{ -\frac{1}{2}(y - \mu(i))^T \Sigma(i)^{-1}(y - \mu(i)) \right\}\,.
\end{equation}

This distribution has the property that continuous variables follow a
multivariate normal distribution $\mathcal{N}_{|\Gamma|}(\mu(i), \Sigma(i))$
conditioned on the discrete variables. The parameters
$\left(p(i), \mu(i), \Sigma(i) \right)$ are called moment characteristics where
$p(i)$ is the probability that $I = i$, and $\mu(i)$ and $\Sigma(i)$ are the
conditional mean and covariance matrix of $Y$ which depend on joint discrete
level $i$. If the covariance matrix is constant across the levels of
$\mathcal{I}$, that is, $\Sigma(i) \equiv \Sigma$, the model is {\it homogeneous}.
Otherwise, the model is said to be {\it heterogeneous}. Throughout this article
we assume an homogeneous mixed GMM underlying the eQTL network. This implies
that genotypes affect only the mean expression level of genes and not the
correlations between them. We can write the logarithm of the density in terms
of the canonical parameters $(g(i), h(i), K(i))$:
\begin{equation}
\log f(i,y) = g(i) + h(i)^Ty - \frac{1}{2}y^TK(i)y\,,\;\;
\label{eq:logdensity}
\end{equation}
where
\begin{eqnarray}
g(i) & = & \log(p(i)) - \frac{1}{2}\log|\Sigma(i)| - \frac{1}{2}\mu(i)^T\Sigma(i)^{-1}\mu(i) - \frac{|\Gamma|}{2}\log(2\pi)\,, \\
h(i) & = & \Sigma(i)^{-1}\mu(i) \,, \label{eq:mixedIntParam} \\
K(i) & = & \Sigma(i)^{-1}\,.
\end{eqnarray}

CG-distributions satisfy the Markov property if and only if their canonical
parameters are expanded into interaction terms such that only interactions
between adjacent vertices are present \citep{Lauritzen:1996}. Thus, \label{pg:K}
\begin{equation}
g(i) = \!\sum_{d\subseteq \Delta} \lambda_{d}(i) \,, \quad 
h_{\gamma}(i) = \!\sum_{d \subseteq \Delta}\eta_{d}(i)_{\gamma}, \quad 
k_{\gamma\eta}(i) = \!\sum_{d \subseteq \Delta}\psi_{d}(i)_{\gamma\eta}\;, 
\label{eq:etas}
\end{equation}
where $\lambda_d(i) > 0$, with $d \subseteq \Delta$ complete in $G$, represent the
discrete interactions among the variables indexed by $d$; $\eta_{d}(i)_\gamma > 0$,
with $d \cup \{\gamma\}$ complete in $G$, represent the mixed interactions
between $X_\gamma$ and the variables indexed by $d$; and
$\psi_{d}(i)_{\gamma\eta} > 0$, with $d \cup \{\gamma, \eta\}$ complete in $G$,
represent the quadratic interactions between $X_\gamma, X_\eta$ and the variables
indexed by $d$. If the model is homogeneous, there are no mixed quadratic
interactions, i.e., $\psi_{d}(i)_{\gamma\eta} = 0$ for $d \neq \emptyset$.
Plugging these expansions in Eq.~(\ref{eq:logdensity}) we obtain
\begin{equation}
\log f(i,y) = \sum_{d \subseteq \Delta} \lambda_{d}(i) + \sum_{d \subseteq \Delta}\sum_{\gamma \in \Gamma} \eta_{d}(i)_\gamma y_\gamma - \frac{1}{2} \sum_{d \subseteq \Delta}\sum_{\gamma, \eta \in \Gamma} \psi_d(i)_{\gamma\eta} y_\gamma y_\eta\,.
\label{eq:logpdf}
\end{equation}

\textbf{Decomposable mixed GMMs:} An important subclass of mixed GMMs is
defined by decomposable marked graphs. A triple $(A,B,C)$ of disjoint subsets
of $V$ form a decomposition of an undirected marked graph $G$ if
$V = A \cup B \cup C$ and: (1) $C$ is a complete subset of $V$; (2) $C$
separates $A$ from $B$; and (3) $C \subseteq \Delta$ or $B \subseteq \Gamma$.
An undirected marked graph $G$ is said to be {\it decomposable} if it is complete,
or if there exists a proper decomposition $(A,B,C)$ such that the subgraphs $G_{A \cup C}$ 
and $G_{B \cup C}$ are decomposable. In different terms, when $G$ is undirected, 
decomposability of $G$ holds if and only if $G$ does not contain chordless cycles 
of length larger than 3 and does not contain any path between two non-adjacent 
discrete vertices passing through continuous vertices only.

\textbf{Maximum Likelihood Estimates of mixed GMMs:}
Let ${\cal X}=\{x^{\left(\nu\right)}\} = \{\left(i^{\left(\nu \right)}, y^{\left(\nu\right)}\right)\}$
be a sample of $\nu = 1,...,n$ independent and identically distributed
observations from a CG-distribution. For an arbitrary subset $A\subseteq V$,
we abbreviate to $i_A=i_{A\cap\Delta}$, $\mathcal{I}_A=\mathcal{I}_{\Delta\cap A}$
and $y_A=y_{A\cap\Gamma}$ and the following sampling statistics are defined:
\begin{eqnarray}
\label{sufstat_n} 
n(i)        \!\! &=&\!\!  \#\left\{\nu: i^{(\nu)}\!= i \right\}\,, \\
\label{sufstat_s}
s(i)        \!\! &=&\!\!  \sum\limits_{\nu: i^{(\nu)} = i} y^{(\nu)}\,, \\ 
\label{sufstat_y}
\bar{y}(i)  \!\! &=&\!\!  s(i)/n(i)\,,\\
\label{sufstat_ss}
ss(i)       \!\! &=&\!\!  \sum\limits_{\nu:i^{(\nu)} = i} y^{(\nu)}(y^{(\nu)})^T\,, \\
ssd(i)      \!\! &=&\!\!  ss(i) - s(i)s(i)^T/n(i)\,, \\
\label{sufstat_ssd}
ssd_A(A)    \!\! &=&\!\!  \sum_{i_A\in\mathcal{I}_A} ssd_{A\cap\Gamma}(i_A)\,. 
\end{eqnarray}

The likelihood function for the homogeneous, saturated model attains its
maximum if and only if $n\geq |\Gamma|+|\mathcal{I}|$, which is almost
surely equal to $n(i)>0$ for all $i\in\mathcal{I}$ \citep[Prop.~6.10]{Lauritzen:1996}.
In such a case the maximum likelihood estimates (MLEs) of the moment
characteristics are defined as
\begin{equation}
\hat{p}(i)=n(i)/n,\quad \hat{\mu}(i)=\bar{y}(i),\quad \hat{\Sigma}=ssd_V(V)/n = ssd/n\,.
\label{eq:momentsHom}
\end{equation}
It follows, then, that saturated mixed GMMs cannot be directly 
estimated from data with $p\gg n$, using only the formulas described above. 
For the unsaturated case, decomposable mixed GMMs also admit explicit MLEs. 
In the homogeneous decomposable case it can be shown \cite[Prop. 6.21]{Lauritzen:1996}
that the MLE exists almost surely if and only if
$n(i_C) \geq |C \cap \Gamma| + |\mathcal{I}_C|$ for all cliques $C$ of $G$ and
$i_C \in \mathcal{I}_C$. In this case, MLEs are defined with the following
canonical parameters \cite[pg. 189]{Lauritzen:1996}: 
\begin{align}
\label{eq:p_decomp}
\hat{p}(i) &= \prod_{j=1}^k \frac{n(i_{C_j})}{n(i_{S_j})}\,, \\
\label{eq:h_decomp}
\hat{h}(i) &= n \Biggl\{ \sum_{j=1}^k \left[ssd_{C_j}(C_j)^{-1}\bar{y}_{C_j}(i_{C_j}) \right]^{|\Gamma|} - \left[ssd_{S_j}(S_j)^{-1}\bar{y}_{S_j}(i_{S_j}) \right]^{|\Gamma|} \Biggr\}\,, \\
\label{eq:K_decomp}
\hat{K}    &= n \Biggl\{ \sum_{j=1}^k\! \left[ssd_{C_j}(C_j)^{-1} \right]^{|\Gamma|}\!\! -\!\! \left[ssd_{S_j}(S_j)^{-1} \right]^{|\Gamma|} \Biggr\}\,,
\end{align}
where $S_1=\emptyset$. The matrices $[M]^{|\Gamma|}$ of Eq.~\eqref{eq:K_decomp}
are defined as follows. Given a matrix $M = \{m_{\gamma\eta}\}_{|A|\times|A|}$
of dimension $|A|\times|A|$ with $A \subseteq \Gamma$, $[M]^{|\Gamma|}$ is a
$|\Gamma|\times|\Gamma|$ matrix such that

\begin{equation*}
  [M]_{\gamma\eta}^{|\Gamma|} =
  \begin{cases}
    m_{\gamma\eta}, & \text{if $\{\gamma,\eta\} \in A$}\,,\\
    0, & \text{otherwise}\,.
  \end{cases}
\end{equation*}
Analogously, in Eq.~\eqref{eq:h_decomp}, $[M]^{|\Gamma|}$ is a $|\Gamma|$-length
vector obtained from a $|A|$-length vector $M$. 

\noindent\newline\textbf{Simulation of eQTL network models with mixed GMMs}
\label{sec:data}

In this subsection, we describe how to simulate eQTL networks with homogeneous
mixed GMMs and data from them. We restrict ourselves to the case of a
backcross, in which each genotype marker can have two different alleles AA and AB.
However, these procedures could be extended to more complex cross models
allowing for other than linear additive effects (codominant model) on the mixed
associations, such as dominance effects.

\textbf{Simulation of eQTL network structures:}
The first step to simulate a GMM consists of simulating its associated graph
$G=(V, E)$ that defines the structure of the eQTL network.
eQTLs and expression profiles are represented in $G$ by discrete vertices $\Delta$
and continuous vertices $\Gamma$, respectively, such that $V=\Delta\cup\Gamma$ 
and $|V|=p$. In the context of genetical genomics data we make the assumption that
discrete genotypes affect gene-expression measurements and not the other way
around. Thus, we consider the underlying graph $G$ as a marked graph where some
edges are directed and represented by arrows and some are undirected.
More concretely, $G$ will have  arrows pointing from discrete vertices to
continuous ones, undirected edges between continuous vertices and no edges
between discrete vertices. From this restriction, it follows that there are no
semi-directed cycles and allows one to interpret these GMMs as {\it chain graphs},
which are graphs formed by undirected subgraphs connected by directed edges
\citep{Lauritzen:1996}.

\textbf{Simulation of parameters of a homogeneous mixed GMM:}
Here we show how to simulate the parameters of the homogeneous CG-distribution
represented by $G$ with given marginal linear correlations of magnitude $\rho$
on the pure continuous (gene-gene) associations and given additive effects of
magnitude $a$ on the mixed (eQTL) ones.

\textit{Covariance matrix.} Given the structure of a graph $G$, a random
covariance matrix $\Sigma$ can be simulated as follows. Let $G_\Gamma\subseteq G$
denote the subgraph of $p_\Gamma=|\Gamma|$ pure continuous vertices and let $\rho$
denote the desired mean marginal correlation between each pair of continuous r.v.'s
$(X_\gamma, X_\eta)$ such that $(\gamma,\eta)\in G_\Gamma$. Let
${\cal S}^+(G_\Gamma)\subset{\cal S}^+$ denote the set of all
$p_\Gamma\times p_\Gamma$ positive definite matrices in ${\cal S}^+$ such that
every matrix $S\in{\cal S}^+(G_\Gamma)$ satisfies that $\{S^{-1}\}_{ij}=0$ whenever
$i\neq j$ and $(i, j)\not\in G_\Gamma$.

We simulate $\Sigma$ such that $\Sigma\in{\cal S}^+(G_\Gamma)$ in two steps.
First, we build an initial positive definite matrix $\tilde{\Sigma}_0 \in {\cal S}^+$ 
and, from this, we build the {\it incomplete matrix} $\Sigma_0$ with elements
$\{\sigma_{ij}^0\}$ if either $i=j$ or $(i, j)\in G_\Gamma$, and the remaining
elements unspecified. Second, we search for a {\it positive completion} of
$\Sigma_0$, which consists of filling up $\Sigma_0$ in such a way that the
resulting $\Sigma\in{\cal S}^+(G_\Gamma)$.

It can be shown \citep{GroneWolkowicz:1984} that the incomplete matrix $\Sigma_0$ 
admits a positive completion and that it is unique. This means that, given $\Sigma_0$, 
we can use algorithms for maximum-likelihood estimation or Bayesian conjugate inference 
\citep{Roverato:2002} as matrix completion algorithms. To this end, we first draw 
$\Sigma_0$ from a Wishart distribution $W_{p_\Gamma}(\Lambda, p_\Gamma)$ with
$\Lambda=D\tilde{\Sigma}_0D, D=\textrm{diag}(\{\sqrt{1/p_\Gamma}\}_{p_\Gamma})$ and
$\tilde{\Sigma}_0=\{\tilde{\Sigma}_{0_{ij}}\}_{p_\Gamma\times p_\Gamma}$ where 
$\tilde{\Sigma}_{0_{ij}}=1$ for $i=j$ and $\tilde{\Sigma}_{0_{ij}}=\rho$ for $i\neq j$. 
It is required that $\Lambda\in{\cal S}^+$ and
this happens if and only if $-1/(p_\Gamma-1)<\rho<1$ \cite[pg.~317]{Seber:2007}.
Finally, we apply the iterative regression procedure introduced by 
\citet[pg.~634]{HastieFriedman:2009} for maximum-likelihood estimation of Gaussian 
GMMs with known structure, as matrix completion algorithm to obtain $\Sigma$ from $\Sigma_0$.

\textit{Probability of discrete levels.} Since we are simulating a backcross,
each discrete r.v. takes two possible values $i_{\delta} = \{1,2\}$ with equal
probability $p(i_{\delta} = 1) = p(i_{\delta} = 2) = 0.5$. For the sole purpose of
simulating eQTL networks we made the simplifying assumption that discrete r.v.'s
representing eQTLs are marginally independent between them. From these two
assumptions it follows that joint levels $i \in \mathcal{I}$ are uniformly
distributed, that is, $p(i) = 1/|\mathcal{I}|$, $\forall i \in \mathcal{I}$. 

\textit{Conditional mean vector.} Let a discrete variable $I_{\delta}$
have an additive effect $a_{\delta\gamma}$ on the continuous variable $Y_{\gamma}$.
In the case of a backcross this implies that:
\begin{equation}
a_{\delta\gamma} = \mu_{\gamma}(1) - \mu_{\gamma}(2) = \frac{1}{|\mathcal{I}|/2} \sum_{i': i_{\delta} = 1} \mu(i') -
                                                       \frac{1}{|\mathcal{I}|/2} \sum_{i': i_{\delta} = 2} \mu(i') \,.
\label{eq:a_eq}
\end{equation} 
Using Eq.~\eqref{eq:mixedIntParam} we can calculate the conditional mean vector
as $\mu(i) = \Sigma \cdot h(i)$. This enables simulating the covariance matrix
independently from discrete r.v.'s while interpreting it as a conditional one to
generate $\mu(i)$. The values of the canonical parameter
$h(i)=\{h_{\gamma}(i)\}_{\gamma\in\Gamma}$ determine the strength of the mixed
interactions between discrete and continuous r.v.'s. Using Eq.~\eqref{eq:etas}
they can be calculated as follows:
\begin{equation*}
h_{\gamma}(i) =
\begin{cases}
	\eta_{\emptyset\gamma}, & \text{if  $(\delta, \gamma) \not\in E$ $\forall \delta \in \Delta$}\,,\\
	\{ \eta_{\delta}(i_{\delta})_{\gamma} \}_{i_{\delta} \in \mathcal{I}_{\delta}} = \{\eta_{\delta}(1)_{\gamma}, \eta_{\delta}(2)_{\gamma}\}, & \text{if $(\delta, \gamma) \in E$, $\delta \in \Delta$}\,.
\end{cases}
\end{equation*}

Without loss of generality, values $\eta_{\emptyset\gamma}$ are set to zero.
To set values $\eta_{\delta}(1)_{\gamma}$ and $\eta_{\delta}(2)_{\gamma}$, so
that both Eqs.~\eqref{eq:mixedIntParam} and \eqref{eq:a_eq} are satisfied, we proceed
as follows. Assume that an eQTL $I_{\delta}$ has a pleiotropic effect on a set of
genes $\{Y_{\gamma}\}_{\gamma \in A_{\delta}}$, where
$A_{\delta} = \{\gamma \in \Gamma : (\delta, \gamma) \in E \}$. By
combining Eqs.~\eqref{eq:mixedIntParam} and \eqref{eq:a_eq}, for each
$\gamma \in A_{\delta}$ we have that
\begin{eqnarray*}
a_{\delta\gamma} &=& \frac{1}{|\mathcal{I}|/2} \sum_{i': i_{\delta} = 1} \sum_{\zeta \in \Gamma} \sigma_{\gamma\zeta}h_{\zeta}(i')
                   - \frac{1}{|\mathcal{I}|/2} \sum_{i': i_{\delta} = 2} \sum_{\zeta \in \Gamma} \sigma_{\gamma\zeta}h_{\zeta}(i') =\\ 
		 &=& \frac{1}{|\mathcal{I}|/2} \sum_{\zeta \in \Gamma} \sigma_{\gamma\zeta}\left\{ \sum_{i': i_{\delta} = 1} h_{\zeta}(i') - \sum_{i': i_{\delta} = 2} h_{\zeta}(i') \right\} \,.
\end{eqnarray*}

Note that if $j, k \in \mathcal{I}$ are two discrete levels such that
$j_{\delta} = 1$, $k_{\delta} = 2$ and
$j_{\Delta \backslash \{\delta\}} = k_{\Delta \backslash \{\delta\}}$, then
$h_{\zeta}(j) = h_{\zeta}(k)$ for all $\zeta \notin A_{\delta}$. It follows that 
for all $\zeta \notin A_{\delta}$ the terms $h_{\zeta}(i')$ in both summations cancel
out and we obtain,
\begin{eqnarray*}
a_{\delta\gamma} &=& \frac{2}{|\mathcal{I}|} \sum_{\zeta \in A_{\delta}} \sigma_{\gamma\zeta}\left\{ \sum_{i': i_{\delta} = 1} h_{\zeta}(i') - \sum_{i': i_{\delta} = 2} h_{\zeta}(i') \right\} =\\
		 &=& \frac{2}{|\mathcal{I}|} \sum_{\zeta \in A_{\delta}} \sigma_{\gamma\zeta}\left\{ \frac{|\mathcal{I}|}{2} \eta_{\delta}(1)_{\gamma} - \frac{|\mathcal{I}|}{2} \eta_{\delta}(2)_{\gamma} \right\} = 
                                             \sum_{\zeta \in A_{\delta}} \sigma_{\gamma\zeta}\left\{ \eta_{\delta}(1)_{\gamma} - \eta_{\delta}(2)_{\gamma} \right\} \,.
\end{eqnarray*}

Let $\eta_{\delta\gamma} =\eta_{\delta}(1)_{\gamma} - \eta_{\delta}(2)_{\gamma}$
and the vectors 
$\boldsymbol{a}_{\delta A_{\delta}} = \{a_{\delta\gamma}\}_{\gamma \in A_{\delta}}$, 
$\boldsymbol{\eta}_{\delta A_{\delta}} = \{\eta_{\delta\gamma}\}_{\gamma \in A_{\delta}}$,
$\boldsymbol{\eta}_{1 A_{\delta}} = \{\eta_{\delta}(1)_{\gamma}\}_{\gamma \in A_{\delta}}$
and 
$\boldsymbol{\eta}_{2 A_{\delta}} = \{\eta_{\delta}(2)_{\gamma}\}_{\gamma \in A_{\delta}}$.
We write the matrix form of the previous expression as,  
\begin{eqnarray}
\boldsymbol{a}_{\delta A_{\delta}} &=& \Sigma_{\{A_{\delta}, A_{\delta}\}}\boldsymbol{\eta}_{\delta A_{\delta}} \,, \nonumber\\ 
\boldsymbol{\eta}_{\delta A_{\delta}} &=& \Sigma_{\{A_{\delta}, A_{\delta}\}}^{-1} \boldsymbol{a}_{\delta A_{\delta}} \,, \nonumber\\
\boldsymbol{\eta}_{1 A_{\delta}} &=& \Sigma_{\{A_{\delta}, A_{\delta}\}}^{-1} \boldsymbol{a}_{\delta A_{\delta}} + \boldsymbol{\eta}_{2 A_{\delta}}\,. \nonumber
\end{eqnarray}

\noindent\newline\textbf{Simulation of eQTL network models of experimental crosses}

We have integrated the algorithms presented above with functions from the
\texttt{R/qtl} package \citep{Broman:2003} to simulate eQTL network models of
experimental crosses and data from them in the following way. First, we simulate
a genetic map with a given number of chromosomes and markers using the
\texttt{sim.map()} function of the \texttt{R/qtl} package.

Second, we simulate a homogeneous mixed GMM in two steps: (a) we define the,
possibly random, underlying regulatory model of eQTLs and gene-gene associations;
(b) we simulate the parameters $(p(i), \mu(i), \Sigma)$ of this homogeneous mixed
GMM according to the procedures described above.

Third, we simulate data from the previous eQTL network model with the function
\texttt{sim.cross()} from the \texttt{R/qtl} package. This function is
overloaded in \texttt{qpgraph} to plug the eQTL associations into the
corresponding genetic loci and return a \texttt{R/qtl} \texttt{cross} object.
The function \texttt{sim.cross()} defined in the \texttt{qpgraph} package
proceeds as follows. First, the genotype data is simulated by the procedures
implemented in the \texttt{R/qtl} package. Genotypes are sampled at each marker
from a Markov chain with transition probabilities that depend on the distance
between markers and a mapping function. eQTLs are placed at the markers and, in
particular, if eQTLs are located sufficiently away from each other, we can
assume that the corresponding discrete r.v.'s are marginally independent between
them. Finally, \texttt{qpgraph} simulates gene expression values according to the
homogeneous mixed GMM by sampling continuous observations from the corresponding
parameters of the CG-distribution
$\mathcal{N}_{|\Gamma|}\left( \mu(i), \Sigma\right)$, given the sampled genotype
$i$ from all joint eQTLs.

\noindent\newline\textbf{Conditional independence tests parametrized by mixed GMMs}

Approaches to learning the structure of a mixed GMM using higher-order
correlations require testing for conditional independence between any two
r.v.'s $X_{\alpha}$ and $X_{\beta}$, such that $\beta \subseteq \Gamma$,  
given a set of conditioning ones $X_Q$, denoted by $\jind{X_{\alpha}}{X_{\beta}} | X_Q$. 
To this end, we use a likelihood-ratio test (LRT) between two models: a saturated 
model $\mathcal{M}_1$, determined  by the complete graph $G^1 = \left(V, E^1 \right)$,
where $V = \{\alpha, \beta, Q \}$ and $E^1 = V \times V$, and a constrained
model $\mathcal{M}_0$, determined by $G^0 = \left(V, E^0 \right)$ with exactly
one missing edge between the two vertices $\alpha, \beta$ and thus
$E^0 = \{V \times V\} \backslash \left(\alpha, \beta\right)$ and 
$Q = V \backslash \{\alpha, \beta\}$. 

Since $G^1$ is complete and $\left(\alpha, \beta, Q\right)$ is a proper decomposition 
of $G^0$, $\mathcal{M}_1$ and $\mathcal{M}_0$ are both decomposable, and therefore, 
they admit explicit MLEs (Eqs.~\ref{eq:p_decomp}-\ref{eq:K_decomp}). Given that
$V = \Delta \cup \Gamma$, we denote by $(\gamma, \zeta)$ a pair of continuous r.v.'s
(i.e., $\gamma, \zeta \in \Gamma$), and by $(\delta, \gamma)$ a pair of mixed r.v.'s
with $\delta \in \Delta$ and, $\gamma \in \Gamma$, so that either
$Q = V \backslash \{\gamma, \zeta\}$ or $Q = V \backslash \{\delta, \gamma\}$ 
are the conditioning subsets. 

In the context of homogeneous mixed GMMs, the null hypothesis of conditional
independence for the pure continuous case, $\jind{\gamma}{\zeta}| Q$, corresponds
to a zero value in the $\left( \gamma, \zeta \right)$ and
$\left( \zeta, \gamma \right)$ entries of the canonical parameter $K$
(see Eqs.~\ref{eq:etas}, \ref{eq:logpdf}). The log-likelihood-ratio statistic,
which is twice the difference of the log-likelihoods of models $\mathcal{M}_0$
and $\mathcal{M}_1$, is reduced to \citep[see][pg. 192]{Lauritzen:1996}:
\begin{equation}
D_{\gamma\zeta.Q} = -2 \ln \left(\frac{\mathcal{L}_0}{\mathcal{L}_1} \right) 
                              = - 2 \ln \left( \frac{|ssd_{\Gamma}||ssd_{\Gamma \backslash \{\gamma,\zeta\}}|}{|ssd_{\Gamma \backslash \{\gamma\}}||ssd_{\Gamma \backslash \{\zeta\}}|} \right)^{n/2} = - 2 \ln \left( \Lambda_{\gamma\zeta.Q} \right)^{n/2}\,.
\label{eq:Dev_cont}
\end{equation}

The null hypothesis of conditional independence in the mixed case,
$\jind{\delta}{\gamma}| Q$, corresponds to an expansion of the canonical
parameter $h_{\gamma}(i)$ where the terms corresponding to $\delta$ are zero,
$\eta_{\delta}(i)_{\gamma} = 0$ (see Eqs.~\ref{eq:etas}, \ref{eq:logpdf}). In
this case, the log-likelihood-ratio statistic is
\citep[pg. 194]{Lauritzen:1996}:
\begin{equation}
D_{\delta\gamma.Q} = - 2 \ln \left( \frac{|ssd_{\Gamma}||ssd_{\Gamma^*}(\Delta^*)|}{|ssd_{\Gamma^*}||ssd_{\Gamma}(\Delta^*)|} \right)^{n/2}
		   = - 2 \ln \left( \Lambda_{\delta\gamma.Q} \right)^{n/2} \,,
\label{eq:Dev_mixed}                               
\end{equation}
where $\Gamma^* = \Gamma \backslash \{\gamma\}$ and $\Delta^* = \Delta \backslash \{\delta\}$. 
Since models $\mathcal{M}_1$ and $\mathcal{M}_0$ are decomposable,
they are collapsible onto the same set of variables $X_{V \backslash \{\gamma\}}$;
see \citep[pg.~86-87]{Edwards:2000} and \citep{Didelez:2004}. This property
implies that the density functions $f$ of $\mathcal{M}_1$ and $\mathcal{M}_0$
can be factorized as
$f_{V} = f_{V\backslash\{\gamma\}}\cdot f_{\gamma|V\backslash\{\gamma\}}$
such that the marginal and conditional densities,
$f_{V\backslash\{\gamma\}} \in \mathcal{M}_{V\backslash\{\gamma\}}$ and
$f_{\gamma|V\backslash\{\gamma\}} \in \mathcal{M}_{\gamma|V\backslash\{\gamma\}}$,
respectively, can be parametrized separately. Therefore, the likelihood function
of $\mathcal{M}_1$  and $\mathcal{M}_0$ can be computed as the product of the
likelihood of the marginal and the conditional models,
$\mathcal{L}_1 = \mathcal{L}_{\gamma|V\backslash\{\gamma\}}^1\cdot\mathcal{L}_{V\backslash\{\gamma\}}^1$
and $\mathcal{L}_0 = \mathcal{L}_{\gamma |V\backslash\{\gamma\}}^0\cdot\mathcal{L}_{V\backslash\{\gamma\}}^0$, respectively.

It follows that the second term of these factorizations corresponds to the same 
saturated model induced by the complete subgraph formed by the vertices in
$V \backslash \{\gamma\}$, $\mathcal{L}_{V\backslash\{\gamma\}}^1 = \mathcal{L}_{V\backslash\{\gamma\}}^0$,
and we have that
\begin{equation}
D_{\gamma\zeta.Q} = -2 \ln \left(\frac{\mathcal{L}_0}{\mathcal{L}_1} \right)  
                                =  -2 \ln \left(\frac{\mathcal{L}_{\gamma |V\backslash\{\gamma\}}^0}{\mathcal{L}_{\gamma |V\backslash\{\gamma\}}^1} \right)
                                = -2 \ln \left(\frac{\hat{\sigma}_{\gamma| V\backslash\{\gamma\}}^0}{\hat{\sigma}_{\gamma| V\backslash\{\gamma\}}^1} \right)^{-n/2}\,,
\label{eq:jointcondmodel}
\end{equation}
where $\hat{\sigma}_{\gamma| V\backslash\{\gamma\}}^0$ and
$\hat{\sigma}_{\gamma| V\backslash\{\gamma\}}^1$ denote the estimates of the
conditional variance of the r.v. $X_{\gamma}$ given the rest of the r.v.'s under
the null and the alternative conditional models
$\mathcal{M}_{\gamma|V\backslash\{\gamma\}}^0$ and
$\mathcal{M}_{\gamma|V\backslash\{\gamma\}}^1$, respectively. In particular,
these conditional models are equivalent to the ANCOVA models
\cite[pg. 91]{Edwards:2000} in which the continuous r.v.
$\gamma \in \Gamma$ is the response variable and the rest are explanatory. 
In this context, we have that
$\hat{\sigma}_{\gamma| V\backslash\{\gamma\}}^0 = \text{RSS}_0/n$ and
$\hat{\sigma}_{\gamma| V\backslash\{\gamma\}}^1 = \text{RSS}_1/n$, where
RSS is the residual sum of squares of the corresponding ANCOVA model
and $n$ is the sample size. The ANCOVA model corresponding to
$\mathcal{M}_{\gamma|V\backslash\{\gamma\}}^1$ for the case of a backcross is,
\begin{equation}
 X_{\gamma} = \mu + \beta_{\delta}Z_{\delta} + \sum_{\kappa\in \Delta^*} \beta_{\kappa} Z_{\kappa} +
              \sum_{\kappa_1=2}^{|\Delta|} \sum_{\kappa_2=1}^{{|\Delta| \choose \kappa_1}} 
              \left[\beta_{\kappa_1\kappa_2} \left(\prod_{\kappa=1}^{\kappa_1} {Z}_{\kappa}\right)\right] + 
              \sum_{\lambda \in \Gamma^*} \beta_{\lambda} X_{\lambda} + \epsilon \,.
\label{eq:ch4:ancova1}
\end{equation}
In this model, $\mu$ is the phenotype's mean, the term $\beta_{\delta}Z_{\delta}$
represents the effect of the discrete variable $\delta \in \Delta$ that we are
testing and we assume that $\epsilon\sim\mathcal{N}\left(0,\sigma_{\gamma}^2\right)$. 
The continuous r.v.'s in $Q$, $\Gamma^*$, are modeled as a linear combination of
r.v.'s (third summation of the equation). On the other hand, the joint levels of
$\delta$ and of the discrete r.v.'s in $Q$, $\Delta$, are encoded through
$(|\mathcal{I}| -1)$ terms where some of them represent the main effects of each
discrete r.v. (first summation). The rest of the terms encode all the interacting
effects between the discrete r.v.'s (second summation). Each variable $Z_{\kappa}$
is an indicator variable that, in the case of a backcross, takes values 0 or 1 if the
genotype of $I_\kappa$ is AA or AB, respectively.

The count of the number of parameters in the model of Eq.~\eqref{eq:ch4:ancova1} is
as follows: $|\Delta|$ parameters come from the term encoding the main effect of
$Z_{\delta}$ and the first summation; the second summation involves
$2^{|\Delta|} - 1 - |\Delta|$ terms whereas the third one involves $|\Gamma| - 1$
terms. Thus, the saturated model $\mathcal{M}_{\gamma|V\backslash\{\gamma\}}^1$
described in Eq.~\eqref{eq:ch4:ancova1} has $2^{|\Delta|}+|\Gamma|-2$ free parameters
in total, and therefore, $n - 2^{|\Delta|} - |\Gamma| + 2$ degrees of freedom,
where $n$ is the sample size of the data.

For the pure continuous case, the conditional model corresponding to
$\mathcal{M}_{\gamma|V\backslash\{\gamma\}}^0$ is the same as the one in
Eq.~\eqref{eq:ch4:ancova1} except that we remove the term of the third summation 
that corresponds to the variable $X_{\zeta}$. In this case, this model has
$n - 2^{|\Delta|} - |\Gamma| + 3$ degrees of freedom. Therefore, under the null
hypothesis, the statistic $D_{\gamma\zeta.Q}$ follows asymptotically a
$\chi_{df}^2$ distribution with $df = 1$ degree of freedom.

In general, we can derive the degrees of freedom for the pure continuous case by
writing explicitly the conditional expectation of $X_\gamma$ given $X_{V\backslash\{\gamma\}}$.
Under $\mathcal{M}_{\gamma|V\backslash\{\gamma\}}^1$ this corresponds to,
\begin{equation}
\textrm{E}\left( X_{\gamma} | \Delta, \Gamma \backslash \{\gamma \}\right) = \alpha(i_\Delta) +
\!\!\!\!\sum_{\lambda \in \Gamma \backslash \{\gamma\}}\!\!\!\beta_{\gamma\lambda|\Gamma\backslash\{\gamma\}} X_{\lambda}\,,
\label{eq:condExpectSat}
\end{equation}
where $\alpha(i_\Delta) = \mu_{\gamma}(i_\Delta)\!-\!\sum_{\lambda \in \Gamma \backslash \{\gamma\}}\!\beta_{\gamma\lambda|\Gamma\backslash\{\gamma\}}\mu_{\lambda}(i_\Delta)$ and $\beta_{\gamma\lambda|\Gamma\backslash\{\gamma\}}$
is the partial regression coefficient that is found through the canonical
parameter $K=\{k_{\gamma\zeta}\}, \forall\gamma,\zeta\in\Gamma$, as 
$\beta_{\gamma\lambda|\Gamma\backslash\{\gamma\}}=-k_{\gamma\lambda}/k_{\gamma\gamma}$ 
\cite[pg.~130]{Lauritzen:1996}. This model has $n - |\mathcal{I}| - |\Gamma| + 1$
degrees of freedom since it has $|\mathcal{I}|$ parameters that come from the first
term in Eq.~(\ref{eq:condExpectSat}) and $|\Gamma| - 1$ from the second term.
On the other hand, the conditional expectation of $X_{\gamma}$ given
$X_{V \backslash \{\gamma\}}$ under $M_{\gamma|V\backslash\{\gamma\}}^0$ is
\begin{equation*}
 \textrm{E}\left( X_{\gamma} | \Delta, \Gamma \backslash \{\gamma, \zeta \} \right) = \alpha(i_\Delta) +
 \!\!\!\!\sum_{\lambda \in \Gamma \backslash \{\gamma, \zeta\}}\!\!\!\beta_{\gamma\lambda|\Gamma\backslash\{\gamma,\zeta\}}X_{\lambda}\,,
 \end{equation*} 
which leads to $n - |\mathcal{I}|- |\Gamma|+ 2$ degrees of freedom. By computing
the difference in the degrees of freedom of both models, we have that
$D_{\gamma\zeta.Q}$ follows asymptotically a $\chi_{df}^2$ distribution with
$df = 1$ degree of freedom.

In the mixed case, the likelihood-ratio statistic $D_{\delta\gamma.Q}$ of
Eq.~\eqref{eq:Dev_mixed} is related to the LOD score used in QTL mapping
\[\text{LOD} = \log_{10}\left( \frac{\mathcal{L}_1}{\mathcal{L}_0}\right)\,,\]
through the following transformation of the LOD score:
\begin{equation}
D_{\delta\gamma.Q} = 2\ln(10)\text{LOD}\,.
\label{eq:LODLambda}
\end{equation}
In fact, since the ratio between $\mathcal{L}_1$ 
and $\mathcal{L}_0$ is equivalent to the ratio between
$\mathcal{L}_{\gamma|V\backslash\{\gamma\}}^1$ and
$\mathcal{L}_{\gamma|V\backslash\{\gamma\}}^0$ we have that 
\begin{equation}
\text{LOD} = \log_{10}\left( \frac{\mathcal{L}_1}{\mathcal{L}_0}\right) = 
               \log_{10}\left( \frac{\mathcal{L}_{\gamma|V\backslash\{\gamma\}}^1}{\mathcal{L}_{\gamma|V\backslash\{\gamma\}}^0} \right)\,.
\label{eq:lodconditionalmodel}
\end{equation}

In this case, the conditional saturated model $\mathcal{M}_{\gamma|V\backslash\{\gamma\}}^1$
for a backcross is the same as the one in Eq.~\eqref{eq:ch4:ancova1}. By
contrast, in the conditional constrained model $\mathcal{M}_{\gamma|V\backslash\{\gamma\}}^0$
we delete all the terms of Eq.~\eqref{eq:ch4:ancova1} that involve the r.v. $X_{\delta}$.
This constrained model has $n-2^{|\Delta|-1}-|\Gamma|+2$ free parameters. Thus, for a
backcross the likelihood-ratio statistic $D_{\delta\gamma.Q}$, and therefore, the
transformation of the LOD score (Eq.~\ref{eq:LODLambda}), follows a $\chi_{df}^2$
distribution with $df = 2^{|\Delta|-1}$ degrees of freedom.

Again, we can derive the degrees of freedom of the $\chi^2_{df}$ distribution of the
general case by looking at the conditional expectation of models
$\mathcal{M}_{\gamma|V\backslash\{\gamma\}}^1$ (Eq.~\ref{eq:condExpectSat}),
and $\mathcal{M}_{\gamma|V\backslash\{\gamma\}}^0$ written as
\begin{equation}
\textrm{E}\left( X_{\gamma} | \Delta \backslash \{\delta\} , \Gamma \backslash \{\gamma \} \right) = \alpha(i_{\Delta \backslash \{\delta\}}) + \!\!\!\!\sum_{\lambda \in \Gamma \backslash \{\gamma\}}\!\!\!\beta_{\gamma\lambda|\Gamma\backslash\{\gamma\}} X_{\lambda}\,.
\label{eq:condExpectCons}
\end{equation}
Here, the first term involves $|\mathcal{I}_{\Delta^*}|$ parameters and the
second $|\Gamma|-1$, so that the constrained model has
$n - |\Gamma|-|\mathcal{I}_{\Delta^*}|+1$ free parameters. Hence, we have that
$D_{\delta\gamma.Q}$, and therefore, the transformed LOD score in
Eq.~\eqref{eq:LODLambda} follows a $\chi_{df}^2$ distribution with
$df = |\mathcal{I}_{\Delta^*}|(|\mathcal{I}_{\delta}| - 1)$ degrees of freedom.

In fact, \citet[pg.~192 to 194]{Lauritzen:1996} observed that, for
decomposable mixed GMMs, the likelihood ratios $\Lambda_{\gamma\zeta.Q}$ 
in Eq.~(\ref{eq:Dev_cont}) and $\Lambda_{\delta\gamma.Q}$ in Eq.~(\ref{eq:Dev_mixed}) 
follow exactly a beta distribution. In order to enable such an exact test 
for homogeneous mixed GMMs, we proceed to derive their corresponding
parameters.

Due to the decomposability and collapsibility of the saturated $\mathcal{M}_1$
and constrained $\mathcal{M}_0$ models, we have seen that the analysis of the
joint densities is equivalent to the study of the univariate conditional
densities of $X_{\gamma}$ given the rest of the variables. Concretely, for the
pure continuous case, the likelihood ratio statistic $\Lambda_{\gamma\zeta.Q}$
is equivalent to the ratio $\text{RSS}_1/\text{RSS}_0$ where $\text{RSS}_0$ and
$\text{RSS}_1$ are the residual sum of squares of the constrained and the
saturated univariate models, respectively, and both follow a $\chi_k^2$
distribution, where $k$ is the number of free parameters of each model. 

Let $\text{RSS}_{1.0}$ denote the difference $\text{RSS}_0 - \text{RSS}_1$.
Following \cite[pg.~166]{Rao:1973}, if a r.v. $X$ follows a $\chi_k^2$ with
$k$ degrees of freedom it also follows a gamma distribution $\Gamma(k/2, 2)$.
Hence,
\begin{equation*}
\text{RSS}_1 \sim \Gamma\left( \frac{n - |\Gamma| - |\mathcal{I}| + 1}{2}, 2 \right),\,\, \text{RSS}_0 \sim \Gamma\left( \frac{ n -  |\Gamma| - |\mathcal{I}| + 2}{2} ,2\right) 
\end{equation*}
and $\text{RSS}_{1.0} \sim \Gamma\left(1/2, 2\right)$. Moreover, if $X$ and $Y$
are two independent r.v.'s such that $X \sim \Gamma\left(k_1, \theta \right)$
and $Y \sim \Gamma\left(k_2, \theta \right)$, then it can be
shown~\cite[pg.~165]{Rao:1973} that
\begin{equation}
\frac{X}{X+Y} \sim B(k_1, k_2)\,,
\label{eq:gamma2beta}
\end{equation}
where $B(k_1, k_2)$ denotes the beta distribution with shape
parameters $k_1$ and $k_2$. Finally, if we let $X = \text{RSS}_1$ and
$Y=\text{RSS}_{1.0}$, it follows that,
\begin{equation*}
\Lambda_{\gamma\zeta.Q} = \frac{\text{RSS}_1}{\text{RSS}_0} \sim B\left( \frac{n -  |\Gamma| - |\mathcal{I}| + 1}{2}, \frac{1}{2} \right)\,.
\end{equation*}
By an argument analogous to the pure continuous case, the likelihood-ratio
statistic raised to the power $2/n$ for the null hypothesis of a missing mixed
edge follows a beta distribution with these parameters:
\begin{equation}
\Lambda_{\delta\gamma.Q} \sim B\left(\frac{n - |\Gamma| - |\mathcal{I}| + 1}{2}, \frac{|\mathcal{I}_{\Delta^*}|(|\mathcal{I}_\delta|-1)}{2} \right)\,.
\label{eq:betamixed}
\end{equation}
Hence, the following transformation of the LOD score,
\[\Lambda_{\delta\gamma.Q} = 10^{-\frac{2}{n}\text{LOD}}\,,\]
follows a beta distribution with parameters given in Eq.~\eqref{eq:betamixed}.
On the other hand, if we let $p_1=k_1/2$ and $p_2=k_2/2$ in Eq.~\eqref{eq:gamma2beta},
it can be shown \citep[pg.~167]{Rao:1973} that $F=(X/k_1)/(Y/k_2) \sim F(k_1, k_2)$.
This means that we can also perform an exact conditional independence test in
terms of the $F$-distribution parametrized with a mixed GMM, by first
calculating
\begin{eqnarray}
F_{\gamma\zeta.Q} & = & \frac{1}{n-|\Gamma|-|\mathcal{I}|+1}\cdot\frac{\Lambda_{\gamma\zeta.Q}}{1-\Lambda_{\gamma\zeta.Q}}\,, \nonumber \\
F_{\delta\gamma.Q} & = & \frac{|\mathcal{I}_{\Delta^*}|(|\mathcal{I}_\delta|-1)}{n-|\Gamma|-|\mathcal{I}|+1}\cdot\frac{\Lambda_{\delta\gamma.Q}}{1-\Lambda_{\delta\gamma.Q}}\,, \nonumber
\end{eqnarray}
for the pure continuous and mixed cases, respectively, and then using the fact
that they follow the F-distribution under the null, specified here below as,
\begin{eqnarray}
F_{\gamma\zeta.Q} & \sim & F(1, n-|\Gamma|-|\mathcal{I}|+1)\,, \nonumber \\
F_{\delta\gamma.Q} & \sim & F(|\mathcal{I}_{\Delta^*}|(|\mathcal{I}_\delta|-1), n-|\Gamma|-|\mathcal{I}|+1)\, . \nonumber
\end{eqnarray}
The proportion, denoted by $\eta^2$, of (phenotypic) variance of $X_{\gamma}$
explained by an eQTL $X_{\delta}$ while controlling for the rest of r.v.'s $
X_{V\backslash\{\gamma\}}$, can be estimated as the difference between the
estimated conditional variances of $X_{\gamma}$ given $X_{V\backslash\{\gamma\}}$
under the saturated and the constrained models, divided by the total variance of
$X_{\gamma}$:
\begin{equation}
\eta^2 = \frac{\text{var}\{\text{E}(X_{\gamma} | X_{V \backslash\{\gamma\}})\} - \text{var}\{\text{E}(X_{\gamma} | X_Q)\}}{\text{var}(X_{\gamma})}\,,
\end{equation}
which, after applying the law of total variance, leads to
\begin{equation}
 \eta^2 = \frac{\hat{\sigma}_{\gamma|V \backslash\{\gamma\}}^0 -\hat{\sigma}_{\gamma|V \backslash\{\gamma\}}^1}{\hat{\sigma}_{\gamma\gamma}}
             = \frac{\text{RSS}_0 - \text{RSS}_1}{(n-1)\cdot \text{var}(X_{\gamma})}\,.
\label{eq:etasquared}
\end{equation}
Note that when $V \backslash \{\gamma\} = \{\delta\}$, that is, $Q = \emptyset$,
the estimated conditional variance under the constrained model,
$\text{RSS}_0/n$, is equal to the unconditional variance of $X_{\gamma}$;
i.e., $\text{RSS}_0/n = \text{var}(X_{\gamma})$. In this case, the proportion of
the phenotypic variance explained by the eQTL reduces
to~\cite[pg.~77]{Broman:2009},
\begin{equation}
 \eta^2 = \frac{\text{RSS}_0 - \text{RSS}_1}{\text{RSS}_0} = 1 - \Lambda_{\delta\gamma.Q}\,.
\end{equation}

\noindent\newline\textbf{$q$-Order correlation graphs}
\label{subsec:limitedOrderCor}

There is a wide spectrum of strategies based on testing for conditional
independence, which can be followed to learn a mixed GMM from data, similarly
as with Gaussian GMMs for pure continuous data; see, e.g.,
\citep{Castelo:2006,KalischBuhlmann:2007}. In the context of estimating eQTL
networks from genetical genomics data the number of genes and genotype markers
$p$ exceeds by far the sample size $n$; i.e., $p\gg n$. This fact precludes
conditioning directly on the rest of the genes and markers $X_{V\backslash\{i,j\}}$
when testing for an eQTL association $(i, j)$ while adjusting for all possible
indirect effects. In other words, we cannot directly test for full-order
conditional independences $\jind{X_i}{X_j}|X_{V\backslash\{i,j\}}$.

We approach this problem using limited-order correlations, an strategy
successfully applied to Gaussian GMMs \citep{Castelo:2006}. It consists of
testing for conditional independences of order $q < (p-2)$, i.e.,
$\jind{X_i}{X_j}|X_Q$ with $|Q|=q$, expecting that many of the indirect
relationships between $i$ and $j$ can be explained by subsets $Q$ of
size $q$. The extent to which this can happen depends on the sparseness of the
underlying network structure $G$ and on the number of available observations $n$.
The mathematical object that results from testing $q$-order correlations is
called a $q$-order correlation graph, or qp-graph \citep{Castelo:2006}, and it
is defined as follows.

Let $P_V$ be a probability distribution which is Markov over an undirected graph
$G=(V, E)$ with $|V|=p$ and an integer $0\leq q\leq (p-2)$. A qp-graph of order
$q$ with respect to $G$ is the undirected graph $G^{(q)}=(V, E^{(q)})$ where
$(i, j)\not\in E^{(q)}$ if and only if there exists a set
$U\subseteq V\backslash\{i,j\}$ with $|U|\leq q$ such that $\jind{X_i}{X_j}|X_U$
holds in $P_V$ \citep{Castelo:2006}.

Assuming there are no additional independence restrictions in $P_V$ than those
in $G$, it can be shown \citep{Castelo:2006} that $G \subseteq G^{(q)}$ in the
sense that every edge that is present in the true underlying network $G$ is also
present in the qp-graph $G^{(q)}$. From this fact it follows that a qp-graph
$G^{(q)}$ approaches $G$ as $q$ grows large, and therefore, $G^{(q)}$ can be
seen as an approximation to $G$; see \citet{Castelo:2006} for further details.

Because separation in undirected marked graphs with mixed discrete and
continuous vertices works the same as in undirected pure graphs with either one
of these two types of vertices, it follows that the definition of qp-graph also
holds for mixed vertices and CG-distributions $P_V$.

\noindent\newline\textbf{Estimation of eQTL networks with qp-graphs}
\label{subsec:NRR}

Instead of directly approaching the problem of inferring the graph structure $G$
of the underlying eQTL network from genetical genomics data with $p\gg n$, we
propose to calculate a qp-graph estimate $\hat{G}^{(q)}$. For this purpose, we
measure the association between two r.v.'s by means of a quantity called the
\textit{non-rejection rate} (NRR), which is defined as follows.

Let $\mathcal{Q}_{ij}^q=\{Q\subseteq V\backslash\{i, j\} : |Q|=q \}$ and let
$T_{ij}^q$ be a binary r.v. associated with the pair of vertices 
$(i, j)$ that takes values from the following three-step procedure: 1)
an element $Q$ is sampled from $\mathcal{Q}_{ij}^q$ according to a (discrete)
uniform distribution; 2) test the null hypothesis of conditional independence
$H_0: \jind{X_i}{X_j}|X_Q$; and 3) if the null hypothesis $H_0$ is rejected then
$T_{ij}^q$ takes value 0, otherwise takes value 1. 

We have that $T_{ij}^q$ follows a Bernoulli distribution and the non-rejection 
rate, denoted as $\nu_{ij}^q$, is defined as its expectancy 
\[\nu_{ij}^q := \textrm{E}[T_{ij}^q]=\Pr(T_{ij}^q=1)\,.\]
The NRR measure was originally developed to learn qp-graphs from pure continuous
data \citep{Castelo:2006}. However, note that by using a suitable test for the
null hypothesis $H_0: \jind{X_i}{X_j}|X_Q$, we can also use the NRR in mixed
data sets such as those produced by genetical genomics experiments.

It can be shown \citep{Castelo:2006} that the theoretical NRR is a function of
the probability $\alpha$ of the type-I error of the test, the mean value
$\beta_{ij}$ of the type-II error of the test for all subsets $Q$, and the
proportion $\pi_{ij}^q$ of subsets $Q$ of size $q$ that separate $i$ and $j$ in
the underlying $G$:
\begin{equation}
\nu_{ij}^q = \beta_{ij}(1 - \pi_{ij}^q) + (1 - \alpha)\pi_{ij}^q\,.
\label{eq:nrrdefinition}
\end{equation}
This expression helps understanding the information conveyed by the NRR in the
following way. If a pair of vertices $(i, j)$ is connected in $G$, then
$\pi_{ij}^q = 0$ and $\nu_{ij}^q = \beta_{ij}$. This means that for associations
present in $G$, the NRR $\nu_{ij}^q$ is 1 minus the statistical power to detect
that association. In such a case, $\nu_{ij}^q$ depends on the strength of the
association between $X_{i}$ and $X_{j}$ over all marginal distributions of size
$(q+2)$. Moreover, note that from Eq.~\eqref{eq:nrrdefinition} it follows that a
$\nu_{ij}^q$ value close to zero implies that both, $\beta_{ij}$ and $\pi_{ij}^q$,
are close to zero. This means that either $(i, j)$ is in $G$ or $q$ is too small.

Analogously, if $\nu_{ij}^q$ is large, then either $\pi_{ij}^q$ or $\beta_{ij}$
are large, and we can conclude that either $(i, j)$ is not present in $G$ or,
otherwise, there is no sufficient statistical power to detect that association.
As the statistical power to reject the null hypothesis $H_0$ depends on $n-q$,
the latter circumstance may be due to an insufficient sample size $n$, a value of
$q$ that is too large, or both. From these observations it follows that a NRR
value $\nu_{ij}^q$ close to zero indicates that $(i,j)\in G^{(q)}$ while a value
close to one points to the contrary, $(i,j)\not\in G^{(q)}$.

The estimation of $\nu_{ij}^q$ for a pair of r.v.'s $(X_{i}, X_{j})$ can be
obtained by testing the conditional independence $\jind{X_i}{X_j}|X_Q$ for every
$Q \in \mathcal{Q}_{ij}^q$. However, the number of subsets $Q$ in
$\mathcal{Q}_{ij}^q$ can be prohibitively large. An effective approach to
addressing this problem \citep{Castelo:2006} consists of calculating an estimate
$\hat{\nu}_{ij}^q$ on the basis of a limited number subsets
$Q \in \mathcal{Q}_{ij}^q$, such as 100, sampled uniformly at random.

We may be interested in explicitly adjusting for confounding factors and other
covariates $\mathcal{C} = \{C_1, C_2, \hdots, C_k\}$. It is straightforward to
incorporate them into a NRR $\nu_{ij.\mathcal{C}}^q$ by sampling subsets $Q$
from
\[\mathcal{Q}_{ij.\mathcal{C}}^q = \{ Q \subseteq \left\{V \backslash \{i, j\} \right\} \cup \mathcal{C} : \mathcal{C}\subseteq Q\;\;\textrm{and}\;\; |Q| = q \}\,.\]
Note that covariates in $\mathcal{C}$ can be known or, in the case of unknown
confounding factors, estimated with algorithms such as SVA
\citep{LeekStorey:2007} or PEER \citep{StegleWinn:2010}. Finally, the qp-graph
estimate $\hat{G}^{(q)}$ of the underlying eQTL network structure $G$ can be
obtained by selecting those edges $(i, j)$ that meet a maximum cutoff value
$\epsilon$:
\[\hat{G}^{(q)} := \{(V, E^{(q)}): (i,j) \in E^{(q)} \Leftrightarrow \nu_{ij}^q < \epsilon \}\,.\]

\section{Results}

\noindent\newline\textbf{Flow of genetic additive effects through gene expression}

\begin{figure}[ht]
\centering
\includegraphics[width=0.75\textwidth]{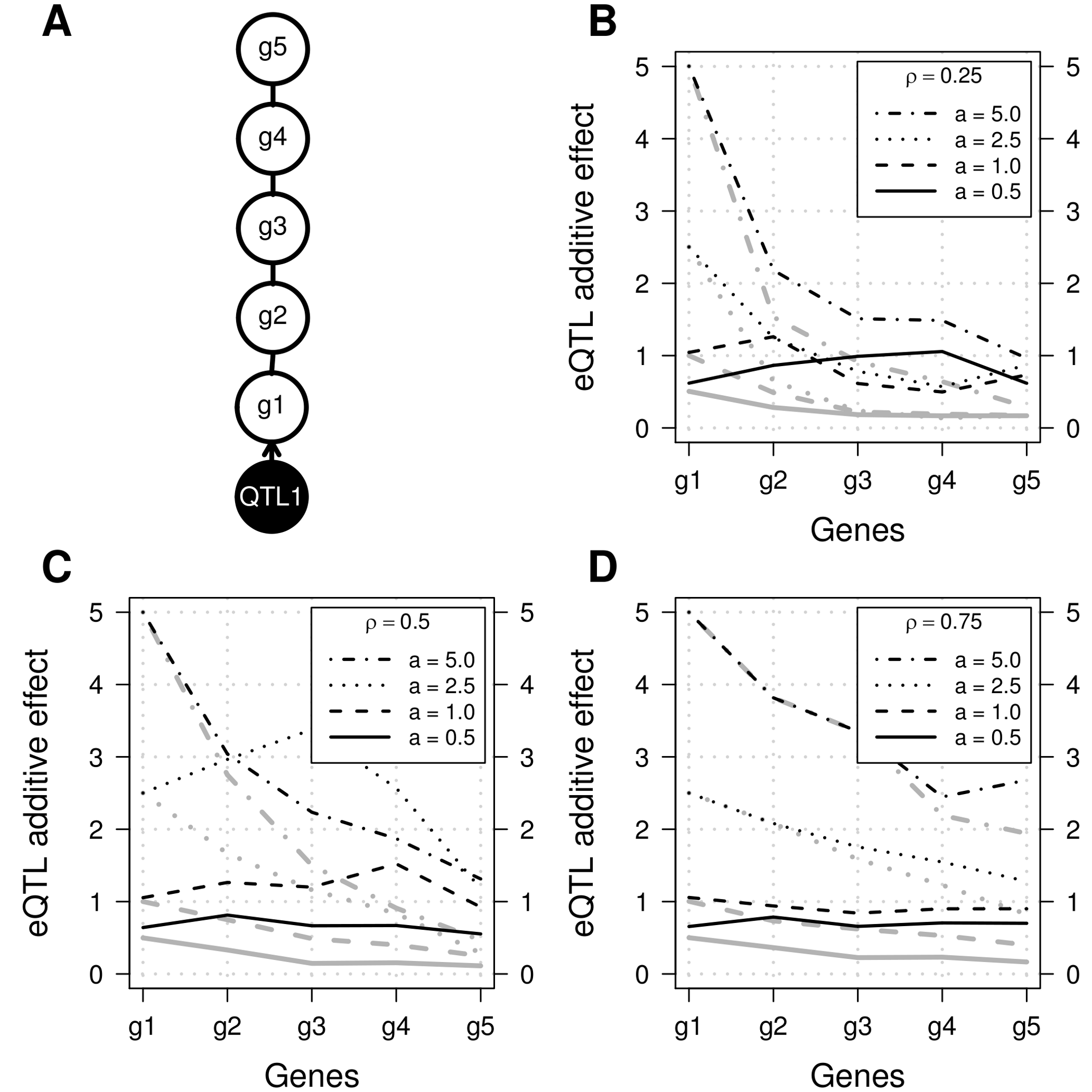}
\caption{Propagation of indirect eQTL additive effects.
(A) Structure of the eQTL network underlying the mixed GMM employed to simulate
the data shown in B-D. (B-D) Average estimated additive effects of the eQTL on
each gene across 10,000 data sets simulated from different combinations of
nominal gene-gene correlations ($\rho$) and additive effects ($a$). Gray lines
were calculated from all data while black lines recreate the effect of selection
bias by using only eQTL associations with LOD $> 3$.}
\label{additiveEffects}
\end{figure}

Using the \texttt{R/qtl} package \citep{Broman:2009} we simulated a genetic map formed
 by one single chromosome 100 cM long and 10 equally-spaced markers. We built
an eQTL network of $p_{\Gamma}=5$ genes forming a chain, where the first of them had one
eQTL placed randomly among the 10 markers (Fig.~\ref{additiveEffects}A). We
simulated 10 mixed GMMs with the eQTL network structure shown in
Figure~\ref{additiveEffects}A, under increasing values of the marginal
correlation between the genes ($\rho=\{0.25, 0.5, 0.75\}$) and of the additive
effect from the eQTL on gene 1 ($a=\{0.5, 1, 2.5, 5\}$). We sampled 1,000 data
sets of $n=100$ observations from each of these 10 models. We estimated the
additive effect of the eQTL on each of the 5 genes, averaged over the 10,000
data sets at each combination of additive effect and marginal correlation. Note
that only the additive effect on gene 1 is direct (Fig.~\ref{additiveEffects}A).

Figure~\ref{additiveEffects}, B-D, gray lines, shows the estimated average
additive effects for the three different marginal correlation values on
the gene-gene associations. These plots demonstrate that additive effects
propagate as a function of gene-gene correlations $\rho$. More concretely, when
$\rho\geq 0.5$ moderate to large additive effects may easily show up as indirect
eQTL associations when inspecting the margin of the data formed by one genetic
variant and one gene-expression profile. Black lines were calculated from the
same data, but discarding additive effects corresponding to LOD scores $\leq 3$.
This recreates the effect of selection bias \citep{Broman:2009} and
shows that indirect genetic additive effects are amplified under this circumstance.

\noindent\newline\textbf{Higher-order conditioning adjusts for confounding effects}

Confounding effects in gene-expression data affect most of the genes being
profiled. Sometimes the sources of confounding are known, or can be estimated
with methods such as SVA \citep{LeekStorey:2007} or PEER \citep{StegleWinn:2010},
and may be explicitly adjusted by including them as main effects into the model.
Often, however, these sources are unknown and it may be difficult to adjust or
remove them without affecting the biological signal and underlying correlation
structure that we want to estimate. Using simulations, here we show that
confounding effects affecting all genes can be implicitly adjusted by using
higher-order conditioning.

Using the same genetic map we simulated before, we built an eQTL network with
100 genes without associations between them and where one of the genes has an
eQTL placed randomly among the 10 markers with a fixed additive effect of $a=2.5$.
A continuous confounding factor was included under two models with $\rho=0.5$:
a systematic one where the confounding factor affects all genes, and a specific
one where it affects only the two genes, or the gene and marker, being tested.
We considered a fixed sample size of $n=100$ and conditioning orders
$q=\{0, 1, \dots, 50\}$, where $q=0$ corresponds to the marginal association
without conditioning. 

We tested the presence of a gene-gene association and of an eQTL association
between the marker containing the simulated eQTL and one of the genes not
associated with that eQTL. Note that none of these associations were present in
the simulated eQTL network. For every $q$ order with $q > 0$, a subset $Q$ of
size $q$ was sampled uniformly at random among the rest of the genes not being
tested, and used for conditioning. When considering the explicit adjustment of
the confounding factor, this one was added to $Q$ except when $q=0$ since then
$Q=\{\emptyset\}$. Once $Q$ was fixed, 100 data sets were sampled from the
corresponding mixed GMM and two conditional independence tests were conducted
in each data set for the presence of both, the eQTL and the gene-gene association,
given the sampled genes in $Q$.

\begin{figure}[ht]
\centering
\includegraphics[width=\textwidth]{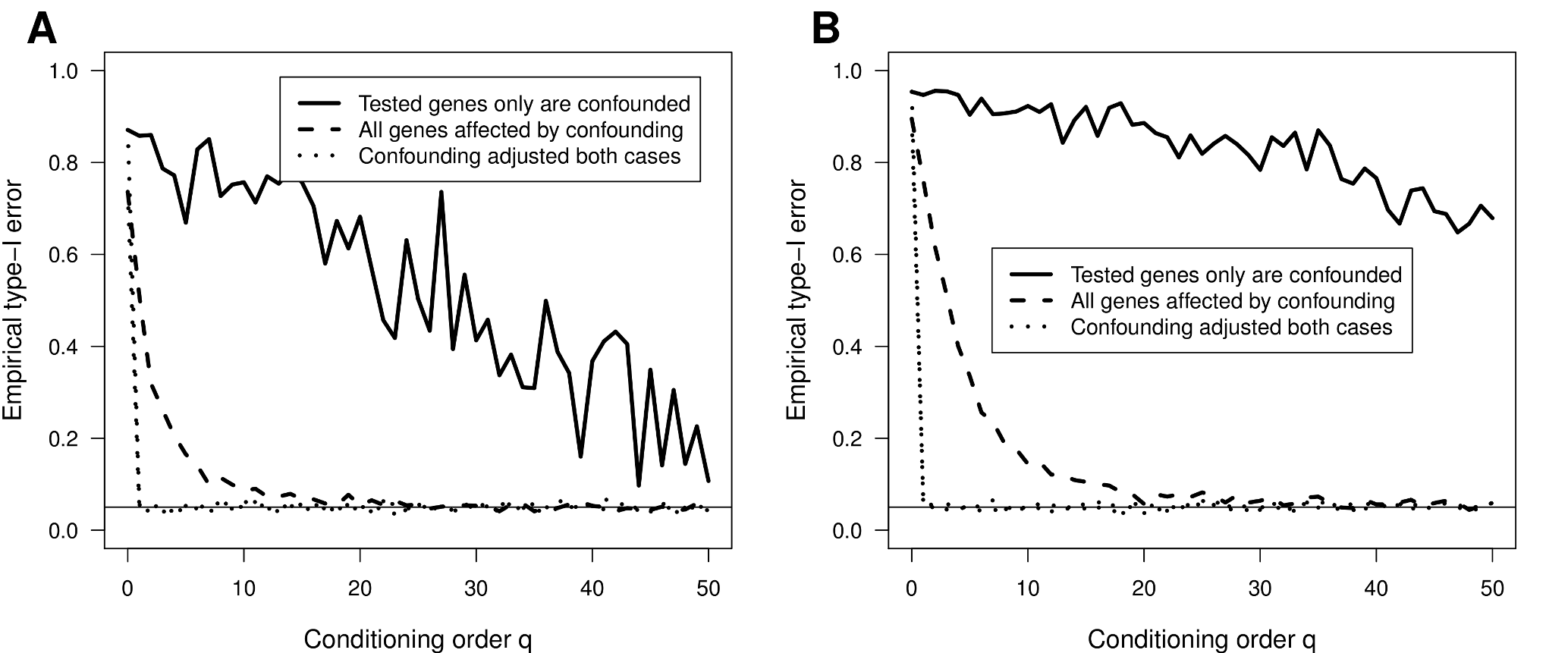}
\caption{Explicit and implicit adjustment of confounding with higher-order
conditional independence tests. Empirical type-I error rate for conditional
independence tests from simulated data at a nominal level $\alpha=0.05$ (dotted
horizontal line) as function of the conditioning order $q$. (A) Results on
testing for an absent eQTL association; (B) results for an absent gene-gene
association. Solid lines correspond to the model under which confounding affects
only the tested genes while dashed lines correspond to a confounding effect on
all genes. Dotted lines from both confounding models overlap because they
correspond to the inclusion of the confounding effect in the conditioning
subsets, thereby explicitly adjusting for it.
}
\label{confoundingEffects}
\end{figure}

Figure~\ref{confoundingEffects} shows the empirical type-I error rate as
function of the conditioning order $q$, where Figure~\ref{confoundingEffects}A
corresponds to the eQTL association and Figure~\ref{confoundingEffects}B to the
gene-gene association. This figure shows that, as expected, the explicit
inclusion of the confounding factor in the conditioning subset $Q$ (dotted
lines) adjusts the confounding effect immediately with $q > 0$ in both
situations, when either all genes are affected or only the tested ones. When
the confounding effect is not included in $Q$ and affects only the tested genes
(solid lines), it yields high type-I error rates that only decrease linearly with
$n-q$, quantity on which statistical power depends. However, when confounding
affects all genes (dashed lines) the type-I error rate has an exponential decay,
and for $q > 20$ the confounding effect is effectively adjusted in these data.

\noindent\newline\textbf{qp-Graph estimates of eQTL networks are enriched for
             {\it cis-}acting associations}
\label{subsec:eQTLnetwork}

Expression QTLs acting in {\it cis} have more direct mechanisms of regulation
than those acting in {\it trans} \citep{RockmanKruglyak:2006,CheungSpielman:2009}.
This hypothesis is supported by the observation that {\it cis}-acting eQTLs
often explain a larger fraction of expression variance and show larger
additive effects, than those acting in {\it trans}
\citep{RockmanKruglyak:2006,PetrettoAitman:2006,CheungSpielman:2009}.
On the other hand, spurious eQTL associations tend to inflate the discovery of
{\it trans}-acting eQTLs \citep{BreitlingJansen:2008}. From this perspective, it
makes sense to expect an enrichment of {\it cis}-eQTLs when indirect
associations are effectively discarded \citep{Kang:2008,Listgarten:2010}.

We NRR values $\nu_{ij}^q$ on every pair $(i, j)$ of marker and gene from the
yeast data set of $n=112$ segregants for different $q=\{25, 50, 75, 100\}$
orders, restricting conditioning subsets to be formed by genes only. The
resulting estimates $\hat{\nu}_{ij}^{q_k}, q_k\in q$, were averaged,
$\hat{\nu}_{ij}^{\bar{q}}=\frac{1}{|q|}\sum_{q_k} \hat{\nu}_{ij}^{q_k}$,
to account for the uncertainty in the choice of the conditioning order $q$
\citep{Castelo:2009}.

We ranked marker-gene pairs $(i, j)$ by average NRR values $\hat{\nu}_{ij}^{\bar{q}}$
and made a comparison against the ranking by $p$-value of the (exact) LRT for
marginal independence (i.e., where $q=0$) to directly assess the added value
of higher-order conditioning under the same type of statistical test. We
considered conservative and liberal cutoff values $\epsilon=\{0.1, 0.5\}$ on the
average NRR $\hat{\nu}_{ij}^{\bar{q}}$ and obtained two different qp-graph
estimates of the underlying eQTL network, denoted by
$\hat{G}_\epsilon^{(\bar{q})}=(V, E_\epsilon^{(\bar{q})})$, each of them having
$|E_{0.1}^{(\bar{q})}|=3,553$ and $|E_{0.5}^{(\bar{q})}|=55,562$ eQTL edges.

\begin{figure}[ht]
\centering
\includegraphics[width=0.7\textwidth]{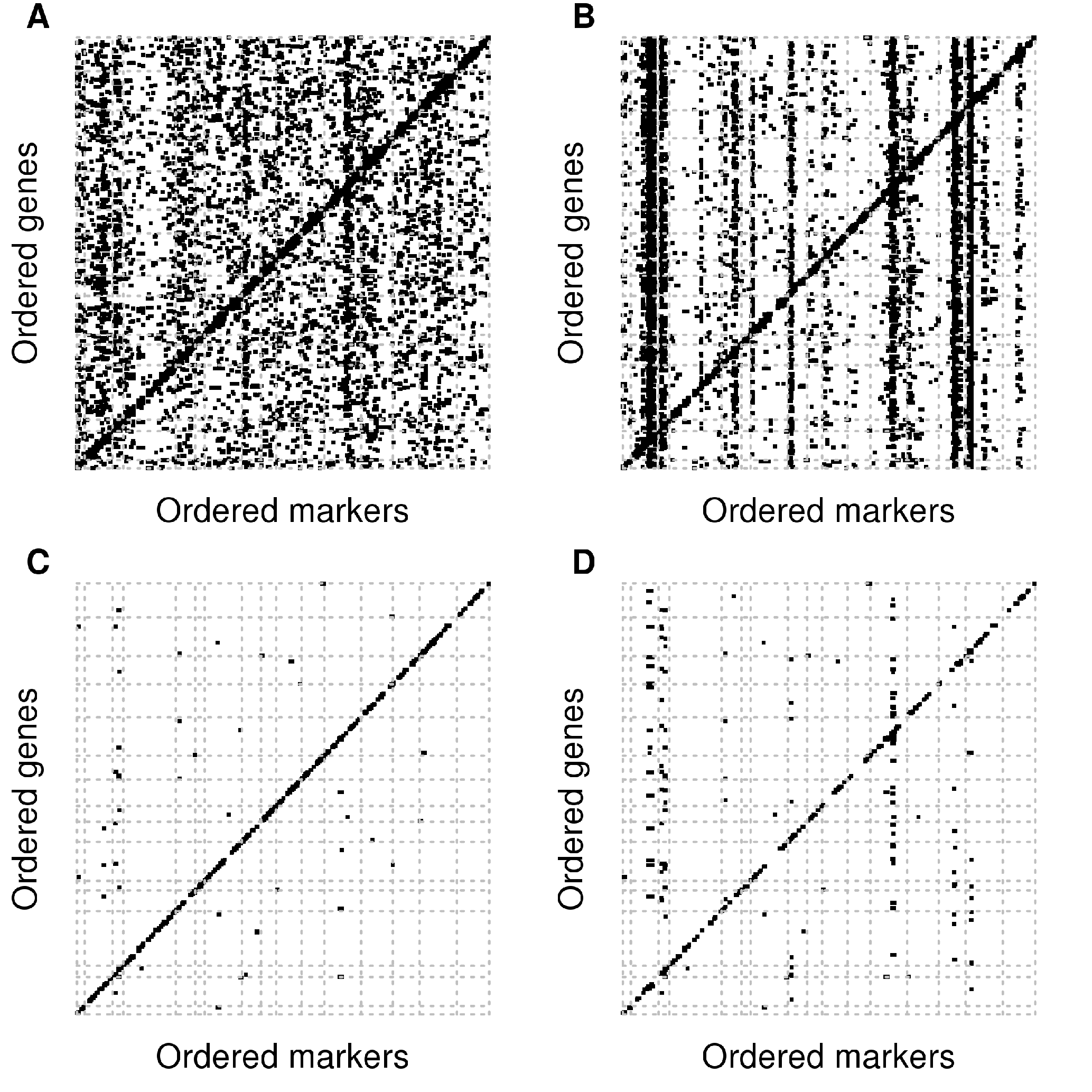}
\caption{Enrichment of {\it cis}-acting eQTL associations.
Dot plots of eQTL associations in yeast, where the $x$-axis and $y$-axis
represent positions along the genome of markers and genes, respectively.
Diagonal bands arise from {\it cis}-eQTLs while vertical ones from
{\it trans}-eQTLs. Each row of two plots shows the top-$k$ eQTLs with
largest strength in terms of non-rejection rates (A and C) and $p$-values
for the null hypothesis of marginal independence (B and D), where $k$ is
the number of eQTLs meeting a liberal (A) and conservative (C) cutoff
on the non-rejection rate. Hence, plots in each row contain the same
number of eQTLs.
}
\label{nrrAvg}
\end{figure}

We then selected the top-$k$ number of marker-gene pairs $(i, j)$ with lowest
$p$-value in the marginal independence test, where
$k=\{|E_\epsilon^{(\bar{q})}|\}$, which led to two other estimates of the eQTL
network, denoted by $\hat{G}_\epsilon^{(0)}$. Note that both,
$\hat{G}_\epsilon^{(\bar{q})}$ and $\hat{G}_\epsilon^{(0)}$, have the same
number of edges, in this case, pairs $(i, j)$ of eQTL associations between a
marker and a gene, thereby enabling a direct comparison of the fraction of
{\it cis}- and {\it trans}-acting selected eQTL associations.

\begin{table}[ht]
\caption{\textbf{Enrichment of {\it cis}-eQTL associations.} Number of
{\it cis}-eQTL associations in yeast found by the method introduced in this
article (\texttt{qpgraph}) and by a marginal test of independence. Different columns
correspond to different cutoffs (conservative, liberal) employed by \texttt{qpgraph}
to select eQTLs, and different distances (500bp and 10Kb) to the gene TSS.}
\begin{center}
\begin{tabular}{|c|c|c|c|c|}\hline
   & \multicolumn{2}{c|}{\parbox[t]{4cm}{\centering {\bf Conservative cutoff} \\ (3,553 eQTLs)}}
   & \multicolumn{2}{c|}{\parbox[t]{4cm}{\centering {\bf Liberal cutoff } \\ (55,562 eQTLs)}} \\ \hline
\parbox[t]{2cm}{\centering \ \\ {\bf Method}} & \parbox[t]{2cm}{\centering {\bf {\it cis} dist. \\ 500bp}}
   & \parbox[t]{2cm}{\centering {\bf {\it cis} dist. \\ 10Kb}}
   & \parbox[t]{2cm}{\centering {\bf {\it cis} dist. \\ 500bp}}
   & \parbox[t]{2cm}{\centering {\bf {\it cis} dist. \\ 10Kb}} \\ \hline
\texttt{qpgraph} & 104 & 1,469 & 369 & 5,878 \\ 
marginal & 56 & 784 & 225 & 3,390 \\ \hline
{\bf Enrichment} & 85\% & 87\% & 64\% & 73\% \\ \hline
\end{tabular}
\end{center}
\label{tab:ciseQTLs}
\end{table}

In Figure~\ref{nrrAvg} we can see dot plots of the eQTL associations present in
qp-graphs $\hat{G}_\epsilon^{(\bar{q})}$ (Figure~\ref{nrrAvg}, A and C), and
those present in $\hat{G}_\epsilon^{(0)}$ using the marginal approach
(Figure~\ref{nrrAvg}, B and D). Given the same number of eQTL associations,
Figure~\ref{nrrAvg} shows that qp-graph estimates of the underlying eQTL network
have a higher number of {\it cis}-acting eQTLs than the marginal test, with an
enrichment between 64\% and 73\% using the liberal cutoff, which increases to
85\% and 87\% using the conservative cutoff (see Table~\ref{tab:ciseQTLs}).
Note that with the marginal approach more vertical bands of {\it trans}-acting
associations remained present among the strongest selected eQTLs
(Figure~\ref{nrrAvg}D), than with the qp-graph estimate (Figure~\ref{nrrAvg}C).
We interpret this observation as evidence of the propagation of additive effects
due to strong gene-gene correlations either present in the underlying eQTL
network or created by confounding effects, and possibly aggravated by selection
bias, as previously shown in Figure~\ref{additiveEffects}.

\noindent\newline\textbf{Performance comparison against another method}
\label{subsec:performanceComparison}

We compared the performance of the methodology introduced in this
paper with a recent approach for causal
inference among pairs of phenotypes \citep{Neto:2013}. This approach, called
causal model selection tests (CMSTs), is implemented in the R/CRAN package
\texttt{qtlhot} and can be used in 3 different ways (parametric, non-parametric
and joint parametric), with two different penalized log-likelihood functions
(AIC and BIC); see \citet[pg.~1005]{Neto:2013} for details.

\begin{figure}[ht]
\centering
\includegraphics[width=\textwidth]{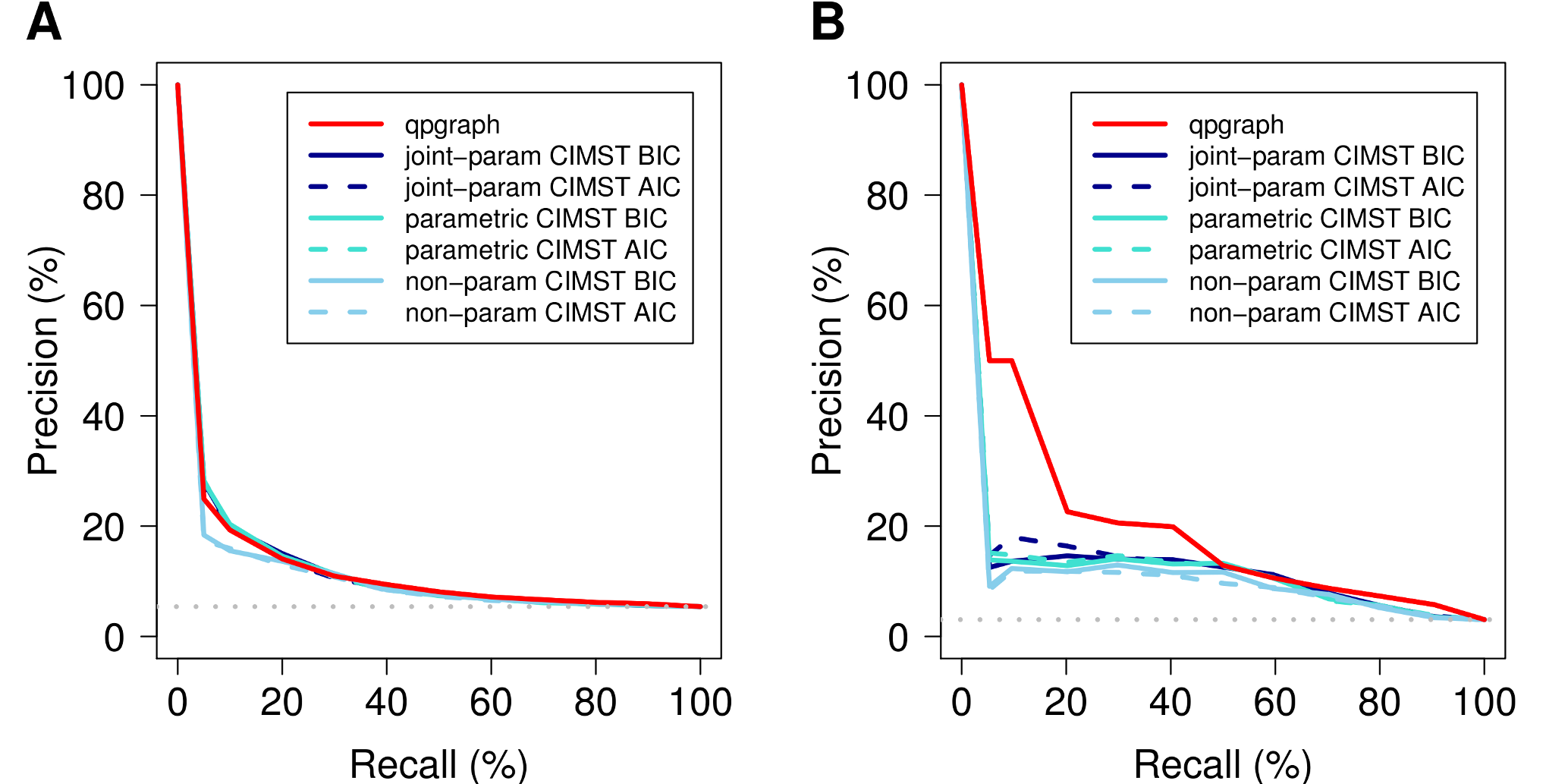}
\caption{Comparison of qpgraph with qtlhot/CMST.
Precision-recall curves calculated from predicted regulatory relationships
inferred from a yeast cross data set using the approach introduced in this paper
(\texttt{qpgraph}) and 6 different configurations of the \texttt{qtlhot}/CMST method.
(A) Predictions are compared against a bronze standard formed by relationships
formed by knocked-out genes and their putative targets derived from differential
expression \citep{Hughes:2000,Zhu:2008}. (B) This bronze standard is further
restricted to relationships also present in the Yeastract database
\citep{Teixeira:2014} of curated transcriptional regulatory associations. The
horizontal gray dotted line indicates the baseline precision attained by a
random predictor.
}
\label{precisionrecall}
\end{figure}

We ran the CMST analysis on the yeast data analyzed in this paper following
the procedure described in \citep[pg.~1008-1010]{Neto:2013}, which assessed
performance using differential expression relationships obtained from a
database of 247 knock-out (KO) experiments in yeast \citep{Hughes:2000,Zhu:2008}.
To enable the comparison we used the raw CMST $p$-value of the so-called
``$M_3$ independent model'' \citep[see][Figure~1]{Neto:2013} to build a ranking of
30,192 potential regulatory relationships, where each pair of genes had a common
eQTL. We compared it against a corresponding
ranking of NRR values estimated from the same data. Among these predicted associations
1,634 formed part of the database of 247 KO-experiments and using them as a
bronze standard, we compared the two rankings by means of precision-recall curves.
These are shown in Figure~\ref{precisionrecall}A and show nearly identical
performance among all compared methods.

Since a KO-gene may produce a cascade of expression changes, many of the 1,634
relationships in the bronze standard may actually be indirect, which would explain
the similar performance using either lower (CMST) or higher (\texttt{qpgraph})
conditioning. We attempted to build a bronze standard of more direct regulatory
relationships by first restricting the initial set of 30,192 possible associations
to 3,074 that involved at least one transcription factor gene. Among these, we found
94 that were present in the database of KO experiments and in the Yeastract database
of curated transcriptional regulatory relationships \citep{Teixeira:2014} and
considered them as new bronze standard for comparison. The resulting precision-recall
curves are shown in Figure~\ref{precisionrecall}B and reveal that NRR values
estimated with \texttt{qpgraph} have larger discriminative power than CMSTs to
identify direct regulatory interactions. Concretely, the area under the curve (AUC) for
\texttt{qpgraph} is between 49\% and 80\% larger than the AUC of the different
versions of CMSTs.

\noindent\newline\textbf{The genetic control of a gene-expression network in a yeast cross}
\label{subsec:eQTLnetworkYeast}

We performed all pairwise exact tests of marginal independence on every marker-gene
and gene-gene pair and corrected the resulting $p$-values by FDR to select those
associations with FDR $< 1\%$. The resulting graph, denoted by $\hat{G}^{(0)}$,
had 92,710 eQTL and 2,203,119 gene-gene associations. The graph $\hat{G}^{(0)}$
constitutes a first estimate of the underlying eQTL network and it could be
also obtained by the classical approach in QTL analysis of permuting phenotypes
to test for the global null hypothesis of no QTL anywhere in the genome. Obviously,
because the associations have been selected only using the margin of the data
formed by one marker and one gene, or two genes, many of them will be indirect or
spurious. To remove those associations we used the previously calculated average NRR
values $\nu_{ij}^{(\bar{q})}$ and selected a conservative cutoff of $\epsilon=0.1$, so that a present
association requires at least 90\% of the conditional independence tests to be
rejected. This resulted in an estimate $\hat{G}_{0.1}^{(\bar{q})}\subseteq\hat{G}^{(0)}$
formed by 3,498 eQTL and 1,799 gene-gene associations. Note that while the estimation
of $\hat{G}^{(0)}$ requires some treatment of multiple testing, the NRR avoids
this problem by actually exploiting the fact that is performing many tests to
inform, by means of a Bernoulli variable, how close or far two genes, or a marker
and a gene, are located in the network.

The genetic connected components of the eQTL network involved 450 genes and
 3,498 eQTLs on 1,493 different loci, with a median of 6 eQTLs per gene. A substantial
percentage of genes (27\%) had $> 10$ eQTLs on their own chromosome. Since eQTLs
in $\hat{G}_{0.1}^{(\bar{q})}$ were independently mapped from each other, a fraction
of those targeting a common gene may be tagging the same causal variant. We removed
redundant eQTLs by the following forward selection procedure. For each gene, we ordered
its linked eQTLs by increasing NRR values $\nu_{ij}^{(\bar{q})}$ and proceeded over
the ranked eQTLs to test the conditional independence of the gene and the eQTL,
given the eQTLs occurring before in the ranking using again the exact test described
in the Methods section. An eQTL association was retained if the test was rejected at
$p<0.05$ and the selection procedure stopped whenever $p > 0.05$ to continue on the next gene.

\begin{figure}[ht]
\centering
\includegraphics[width=\textwidth]{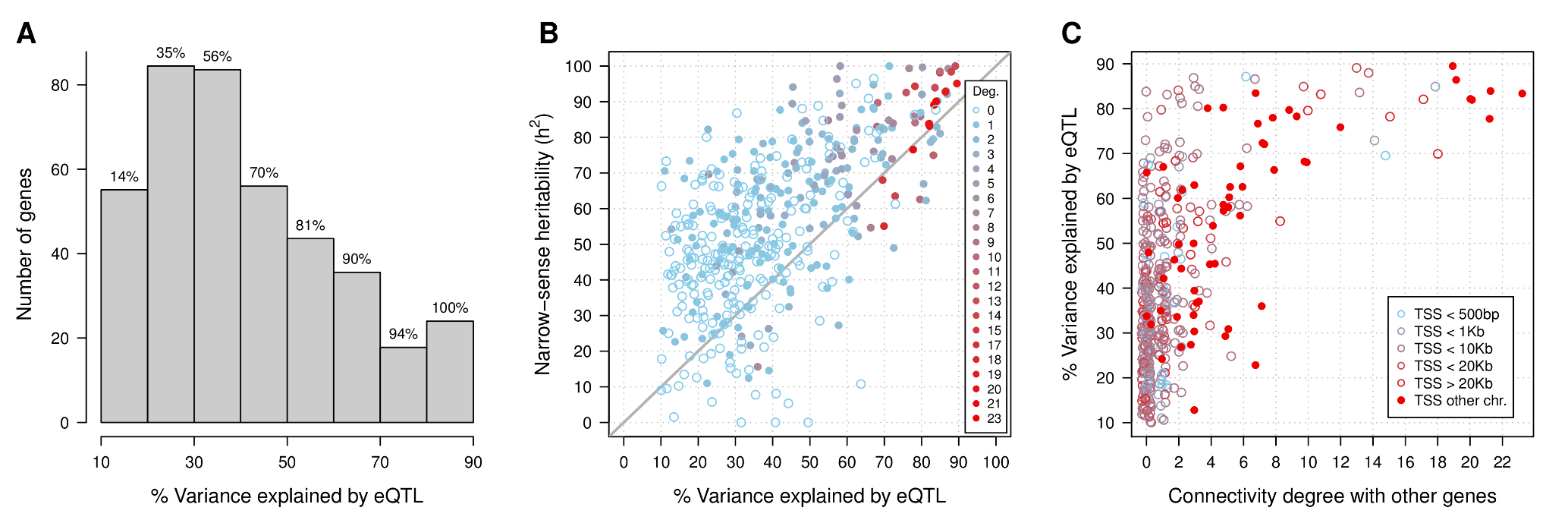}
\caption{Variance explained in the eQTL network.
(A) Distribution of the percentage of gene expression variance explained by
eQTLs. The cumulative percentage of genes is reported on top of each bar.
(B) Scatter plot of the narrow-sense heritability $h^2$ as function of the
percentage of variance explained by eQTLs. The diagonal line is drawn at
values where this percentage equals $h^2$, and it is only shown as a visual
guide. Open circles correspond to genes with exclusively eQTL associations
while solid ones indicate also the presence of at least one association with
other gene. The legend contains the color key for the degree of association
in the network. (C) Percentage of variance explained by eQTLs as function of the
degree of association with other genes in the network. The legend specifies
the color key for the average distance from the transcription start site (TSS)
of the gene to its eQTLs.
}
\label{varexplained}
\end{figure}

The genetic connected components were substantially pruned and the vast majority
of genes (402/450) were left with just one eQTL, 46 genes with two, and only two
had 3 eQTLs. This final eQTL network comprised 500 eQTLs and 1,391 genes,
the vast majority of them (941) forming gene-gene associations without any eQTL.

\textbf{Hub genes have {\it trans-}acting eQTLs with large genetic effects:}
For each of the 450 genes having at least one eQTL, we calculated the percentage
of variance explained by eQTLs using Eq.~\eqref{eq:etasquared}. The distribution of
resulting values is shown in Fig.~\ref{varexplained}A. About 70\% of the genes had
eQTLs explaining 50\% or less of their expression variability, and only in about
10\% of them their eQTLs explained $> 70$\%.

Using a method based on linear mixed modeling and exploiting the relatedness
matrix built from all pairs of segregants
\citep{LeeVisscher:2011,BloomKruglyak:2013} we estimated the narrow-sense
heritability $h^2$ for these 450 genes and compared it against the percentage of
variance explained by the eQTLs (Fig.~\ref{varexplained}B). Setting the
percentage explained to the expected maximum $h^2$ when the former was larger,
the fraction of missing heritability ranges from 0\% to 62\%. This figure also shows
that the connectivity degree to other genes correlates positively with both,
$h^2$ and percentage of variance explained. In fact, as Figure~\ref{varexplained}C
shows, there are 24 genes connected to 9 or more other genes, for which all their
eQTLs explain 68\% or more of their expression variability. Interestingly, this
figure also reveals that half of them have a {\it trans}-acting eQTL located in
a different chromosome. In fact, these particular {\it trans}-eQTLs map all of them
to chromosomes~II and III (Table~\ref{tabGenesDegGe9}). Concretely, the eQTL of gene
{\it SCW11} is located in position 553,857 of chromosome II, 2.6Kb away from the
{\it AMN1} gene. This gene carries a loss-of-function mutation in the BY strain
\citep{Yvert:2003} affecting the expression of daughter-cell-specific genes. The
eQTLs in chromosome III map to the MAT and LEU2 loci, the latter being one of
the engineered deletions in the BY strain.

\begin{table}[ht]
\centering
\caption{\textbf{Hub genes connected to 9 or more other genes in the eQTL network.}
The column ``Pathway'' specifies the primary pathway or molecular process in which the gene is
involved, where (*) indicates that the gene has unknown function and its pathway has been
 predicted using the eQTL network. The column ``Location'' reports the chromosomes where
the gene's eQTLs are located. When both gene and eQTLs are in the same chromosome, the
column distance reports the average distance between them. Columns $h^2$ and $\eta^2$
report, respectively, the narrow-sense heritability and the fraction of variance
explained by the eQTL. The column ``Deg.'' (degree) gives the number of associated genes
in the network.}
\bigskip
{\small
\begin{tabular}{lclcrrrr}
  \hline
{\bf Gene} & {\bf Chr} & {\bf Pathway} & {\bf Location} & {\bf Distance} & $h^2$ & $\eta^2$ & {\bf Deg.} \\ 
  \hline
  STE2 & VI & Mating reg. & III & NA & 0.89 & 0.83 & 23 \\ 
  YKL177W & XI & Mating reg. & III & NA & 0.77 & 0.78 & 21 \\ 
  STE3 & XI & Mating reg. & III & NA & 0.90 & 0.84 & 21 \\ 
  BAR1 & IX & Mating reg. & III & NA & 0.83 & 0.82 & 20 \\ 
  MF(ALPHA)1 & XVI & Mating reg. & III & NA & 0.84 & 0.82 & 20 \\ 
  STE6 & XI & Mating reg. & III & NA & 0.95 & 0.90 & 19 \\ 
  AFB1 & XII & Mating reg. & III & NA & 0.93 & 0.86 & 19 \\ 
  HMLALPHA2 & III & Mating reg. & III & 188156 & 0.55 & 0.70 & 18 \\ 
  MATALPHA1 & III & Mating reg. & III & 732 & 0.98 & 0.85 & 18 \\ 
  HMLALPHA1 & III & Mating reg. & III & 187892 & 0.84 & 0.82 & 17 \\ 
  YCL065W & III & Mating reg. (*) & III & 187423 & 0.94 & 0.78 & 15 \\ 
  YCR041W & III & Mating reg. (*) & III & 263 & 0.68 & 0.70 & 15 \\ 
  MATALPHA2 & III & Mating reg. & III & 996 & 0.63 & 0.73 & 14 \\ 
  ASP3-3 & XII & Nitr. starvation & XII & 14538 & 0.98 & 0.88 & 14 \\ 
  ASP3-1 & XII & Nitr. starvation & XII & 10806 & 1.00 & 0.89 & 13 \\ 
  ASP3-2 & XII & Nitr. starvation & XII & 5046 & 0.94 & 0.84 & 13 \\ 
  LEU1 & VII & Leu biosynthesis & III & NA & 0.93 & 0.76 & 12 \\ 
  YCR097W-A & III & Mating reg. (*) & III & 105944 & 0.75 & 0.83 & 11 \\ 
  HMRA1 & III & Mating reg. & III & 88278 & 0.63 & 0.80 & 10 \\ 
  BAT1 & VIII & Leu biosynthesis & III & NA & 0.83 & 0.68 & 10 \\ 
  OAC1 & XI & Leu biosynthesis & III & NA & 0.90 & 0.68 & 10 \\ 
  ASP3-4 & XII & Nitr. starvation & XII & 18190 & 0.98 & 0.85 & 10 \\ 
  SCW11 & VII & Daugh. cell sep. & II & NA & 0.86 & 0.80 & 9 \\ 
  MFA2 & XIV & Mating reg. & III & NA & 0.84 & 0.78 & 9 \\ 
   \hline
\end{tabular}
}
\label{tabGenesDegGe9}
\end{table}

Genes {\it YCL065W}, {\it YCR041W} and {\it YCR097W-A} in Table~\ref{tabGenesDegGe9}
have currently unknown function. However, a Gene Ontology (GO) enrichment analysis
on each subset of genes connected to them in the eQTL network showed that they are
potentially involved in mating-specific regulatory processes (Holm's FWER $< 0.05$).

Therefore, these highly-connected genes, shown in Table~\ref{tabGenesDegGe9}, are
involved in regulatory processes related to mating-specific expression, reacting
upon nitrogen starvation, participation in the leucine biosynthesis pathway and
daughter cell separation. These are fundamental pathways for yeast growth and
render these results consistent with previous evidence from yeast genetic
interaction networks derived from double-mutant screens
\citep{CostanzoBoone:2010}, where highly-connected genes were involved in
primary cellular functions \citep{BaryshnikovaBoone:2013}.

\begin{figure}[ht]
\centering
\includegraphics[width=0.8\textwidth]{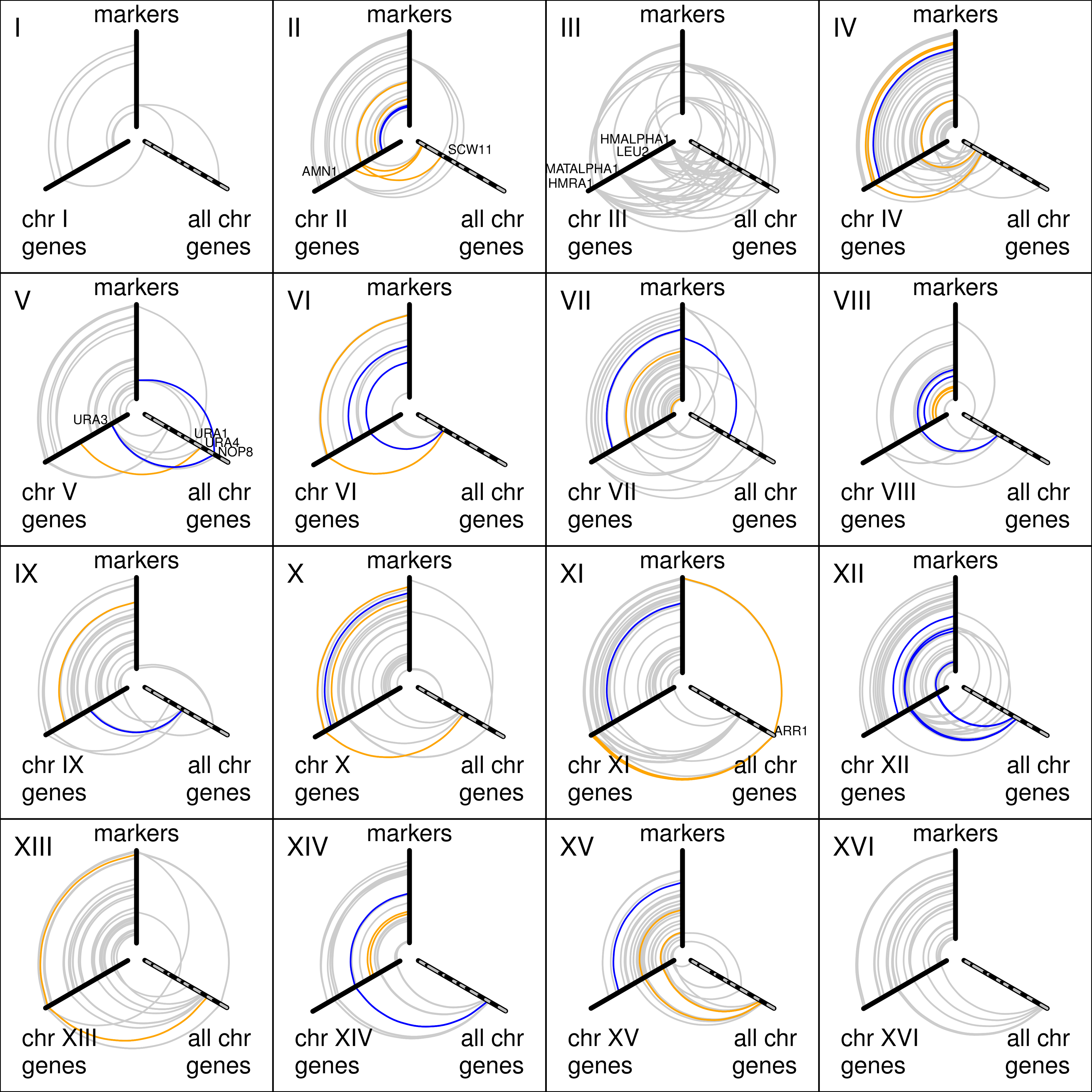}
\caption{eQTL network of a yeast cross.
Hive plots of an eQTL network estimated from a yeast cross, involving only
connected components with at least one eQTL association. For each
chromosome, the hive plot shows three axes, where markers and genes are ordered
from the center according to their genomic location. Vertical and left axes
represent the chromosome in the corresponding plot, while the right axis
represents the entire yeast genome alternating black and gray along consecutive
chromosomes. Edges between genes axes correspond to gene-gene associations.
Edges where at least one of their endpoints corresponds to a transcription factor
or RNA-binding coding gene, are highlighted in orange and blue, respectively.
}
\label{hivePlot}
\end{figure}

\textbf{Differential genetic control of gene expression across chromosomes:}
We also investigated how genetic variation affects gene expression differently
across the yeast chromosomes. For this purpose, we produced hive plots
\citep{Krzywinski:2012} shown in Figure~\ref{hivePlot}, using the R/CRAN package
HiveR \citep{Hanson:2014}. A first observation is that eQTLs occurring within
the same chromosome (edges between the markers axis and genes axis of the same
chromosome) mostly lead to concentric edges, pointing to {\it cis}-regulatory
mechanisms acting at different distances. A remarkable exception is
chromosome~III where many of those edges cross through each other. This
chromosome is also distinctive in that it has a lower density of {\it cis}-acting
eQTLs than the rest of the genome.

Between 81Kb and 92Kb from the beginning of chromosome~III there is a
{\it cis}-acting eQTL on genes \textit{LEU2} and \textit{NFS1}. This eQTL is also
\textit{trans}-associated with genes \textit{BAT1}, \textit{OAC1}, \textit{LEU1}
and \textit{BAP2} located in different chromosomes and involved in the leucine
biosynthesis pathway. According to the UCSC Genome Browser
(\url{http://genome.ucsc.edu}) they all have upstream a binding site of
\textit{LEU3}, a major regulatory switch in this pathway. The engineered deletion
of \textit{LEU2} affects \textit{LEU3} in a feedback loop and this would lead to
expression changes in its targets \citep{Chin:2008}.

Downstream, at about 200Kb of the beginning of chromosome~III, we find the
\textit{MAT} locus whose genetic composition determines the mating type of
yeast. This eQTL is \textit{cis}-associated with the gene \textit{MATALPHA1} which
is expressed in haploids of the alpha mating type and which has been previously
reported as a candidate regulator of the rest of genes associated with this locus
\citep{Yvert:2003, Curtis:2013}. This locus is \textit{trans}-associated with two
other genes in the same chromosome (\textit{HMLALPHA1}, \textit{HMRA1}) and to
a set of genes distributed throughout the genome (\textit{STE2}, {\it STE3},
\textit{STE6}, \textit{AFB1}, \textit{BAR1}, \textit{MF(ALPHA)1}, \textit{MFA2})
which are all involved in the regulation of mating-type specific transcription.

In chromosome~V, around the locus of \textit{URA3} we find a {\it cis}-acting
eQTL which has a {\it trans}-acting effect on \textit{URA1} (chrom.~XI) and
\textit{URA4} (chrom.~XII), the three of them taking part in the
biosynthesis of pyrimidines \citep{Yvert:2003, Curtis:2013}. As it can be easily
seen from Figure~\ref{hivePlot}, and consistent with previous observations
\citep{Yvert:2003}, few of the eQTLs affect directly transcription factors, such
as the {\it ARR1} gene in chromosome~XI, or RNA-binding proteins, such as
{\it NOP8} in chromosome~V.

The edges between the genes axes, correspond to gene-gene associations from the
corresponding chromosome to the rest of the genome and where at least one of the
genes has an eQTL. This is a fraction (503) of a total of 1,799 gene-gene
associations on the entire eQTL network, more directly affected by the genetic
control of gene expression. We observed a systematic pattern of association
between genes from the same chromosome (see, e.g., chrom.~XVI). Replacing the
axis displaying genes from all chromosomes by another gene axis again of the same
chromosome (data not shown) reveals that a fraction of the gene-gene associations
in which one of the genes has a {\it cis}-acting eQTL, occur between genes close
to each other on the chromosome. It may be possible that inherited coexpression
segregates due to linkage disequilibrium and/or that tandem gene duplication
events render genes close to each other being coexpressed. Using the strategies
introduced in this paper further, such as conditioning these associations on
nearby genes, may help to elucidate what fraction of them are of genetic, molecular
or evolutionary origin.

\section*{Discussion}

Gene expression is a high-dimensional multivariate trait whose variability is
the result of genetic, molecular and environmental perturbations, and often
different kinds of confounding effects. Dissecting the components of this
variability and being able to adjust for some of them is a major goal in every
study using genetical genomics data. Here we have used a class of statistical
models with a graphical interpretation, mixed GMMs, to approach this
challenge from a multivariate perspective. Using simulations we have shown
that genetic effects can propagate proportionally to the marginal correlation
between the genes, and that this effect may be amplified under selection bias
(Fig.~\ref{additiveEffects}), underscoring the need to adjust for indirect
associations.

Using standard linear theory and basic principles from mixed GMMs, we have
derived the parameters, in terms of a mixed GMM, for an exact likelihood-ratio
test (LRT) on data from conditional Gaussian distributions that accommodate
both linear and interaction effects between genetic variants and continuous
gene-expression profiles. Higher-order conditioning on mixed data unlocks a
number of strategies that one may follow to disentangle direct and indirect
effects in genetical genomics experiments. We exploited it by using marginal
distributions and $q$-order correlation graphs. We showed that this approach
allows one to adjust for confounding effects (Fig.~\ref{confoundingEffects}),
increases our power to identify \textit{cis}-acting eQTLs (Fig.~\ref{nrrAvg})
and direct regulatory relationships (Fig.~\ref{precisionrecall}), and helped
comparing the genetic control of gene expression between chromosomes
(Fig.~\ref{hivePlot}) and throughout the gene network (Fig.~\ref{varexplained}).
In particular, we could see that the degree of connections of each eQTL gene to
other genes in the eQTL network correlated positively with both, narrow-sense
heritability and fraction of variance explained by eQTLs (Fig.~\ref{varexplained}).
An important fraction of large genetic effects were in fact due to
{\it trans}-acting eQTLs from those genes with more connections. Using those
connections we could predict potential pathways involving three such hub genes
of unknown function. Genetical genomics data from large experimental crosses are
becoming increasingly available to the community. We believe that mixed GMMs can
play a crucial role in harnessing these data to explore linear and interacting
eQTL associations of arbitrary order and advance our understanding of the
genetic control of gene expression.

\section*{Acknowledgments}

This work has been supported by a grant from the Spanish Ministry of Economy and
Competitiveness to R.C. (ref. TIN2011-22826). We thank M.M. Alb\`a, D.R. Cox,
M. Francesconi, S.L. Lauritzen, B. Lehner and N. Wermuth for helpful
discussions on different parts of this paper, and the two anonymous reviewers
for their valuable comments that have improved this paper. We thank R. Brem for
kindly providing raw expression data files from the yeast data set analyzed in
this paper.

\bibliography{eQTLmixedGMMs}

\begin{thebibliography}{60}
\expandafter\ifx\csname natexlab\endcsname\relax\def\natexlab#1{#1}\fi

\bibitem[{{\sc Baryshnikova} {\em et~al.\/}(2013){\sc Baryshnikova}, {\sc
  Costanzo}, {\sc Myers}, {\sc Andrews} and {\sc
  Boone}}]{BaryshnikovaBoone:2013}
{\sc Baryshnikova, A.}, {\sc M.~Costanzo}, {\sc C.~L. Myers}, {\sc B.~Andrews},
  and {\sc C.~Boone}, 2013 Genetic interaction networks: Toward an
  understanding of heritability.
\newblock Annu Rev Genom Hum G {\bf 14}: 111--133.

\bibitem[{{\sc Bing} and {\sc Hoeschele}(2005)}]{Bing:2005}
{\sc Bing, N.}, and {\sc I.~Hoeschele}, 2005 Genetical genomics analysis of a
  yeast segregant population for transcription network inference.
\newblock Genetics {\bf 170}: 533--42.

\bibitem[{{\sc Bloom} {\em et~al.\/}(2013){\sc Bloom}, {\sc Ehrenreich}, {\sc
  Loo}, {\sc Lite} and {\sc Kruglyak}}]{BloomKruglyak:2013}
{\sc Bloom, J.~S.}, {\sc I.~M. Ehrenreich}, {\sc W.~T. Loo}, {\sc T.-L.~V.
  Lite}, and {\sc L.~Kruglyak}, 2013 Finding the sources of missing
  heritability in a yeast cross.
\newblock Nature {\bf 494}: 234--237.

\bibitem[{{\sc Breitling} {\em et~al.\/}(2008){\sc Breitling}, {\sc Li}, {\sc
  Tesson}, {\sc Fu}, {\sc Wu} {\em et~al.\/}}]{BreitlingJansen:2008}
{\sc Breitling, R.}, {\sc Y.~Li}, {\sc B.~M. Tesson}, {\sc J.~Fu}, {\sc C.~Wu},
  {\em et~al.\/}, 2008 Genetical genomics: spotlight on {QTL} hotspots.
\newblock PLoS Genet {\bf 4}: e1000232.

\bibitem[{{\sc Brem} and {\sc Kruglyak}(2005)}]{Brem:2005}
{\sc Brem, R.~B.}, and {\sc L.~Kruglyak}, 2005 The landscape of genetic
  complexity across 5,700 gene expression traits in yeast.
\newblock P Natl Acad Sci USA {\bf 102}: 1572--7.

\bibitem[{{\sc Brem} {\em et~al.\/}(2002){\sc Brem}, {\sc Yvert}, {\sc Clinton}
  and {\sc Kruglyak}}]{BremKruglyak:2002}
{\sc Brem, R.~B.}, {\sc G.~Yvert}, {\sc R.~Clinton}, and {\sc L.~Kruglyak},
  2002 Genetic dissection of transcriptional regulation in budding yeast.
\newblock Science {\bf 296}: 752--755.

\bibitem[{{\sc Broman} and {\sc Sen}(2009)}]{Broman:2009}
{\sc Broman, K.~W.}, and {\sc S.~Sen}, 2009 {\em A guide to {QTL} mapping with
  {R}/qtl\/}.
\newblock Springer.

\bibitem[{{\sc Broman} {\em et~al.\/}(2003){\sc Broman}, {\sc Wu}, {\sc Sen}
  and {\sc Churchill}}]{Broman:2003}
{\sc Broman, K.~W.}, {\sc H.~Wu}, {\sc S.~Sen}, and {\sc G.~A. Churchill}, 2003
  {R}/qtl: {QTL} mapping in experimental crosses.
\newblock Bioinformatics {\bf 19}: 889--890.

\bibitem[{{\sc Castelo} and {\sc Roverato}(2006)}]{Castelo:2006}
{\sc Castelo, R.}, and {\sc A.~Roverato}, 2006 A robust procedure for
  {G}aussian graphical model search from microarray data with p larger than n.
\newblock J Mach Learn Res {\bf 7}: 2621--50.

\bibitem[{{\sc Castelo} and {\sc Roverato}(2009)}]{Castelo:2009}
{\sc Castelo, R.}, and {\sc A.~Roverato}, 2009 Reverse engineering molecular
  regulatory networks from microarray data with qp-graphs.
\newblock J Comput Biol {\bf 16}: 213--227.

\bibitem[{{\sc Chaibub~Neto} {\em et~al.\/}(2013){\sc Chaibub~Neto}, {\sc
  Broman}, {\sc Keller}, {\sc Attie}, {\sc Zhang} {\em et~al.\/}}]{Neto:2013}
{\sc Chaibub~Neto, E.}, {\sc A.~T. Broman}, {\sc M.~P. Keller}, {\sc A.~D.
  Attie}, {\sc B.~Zhang}, {\em et~al.\/}, 2013 Modeling causality for pairs of
  phenotypes in systems genetics.
\newblock Genetics {\bf 193}: 1003--1013.

\bibitem[{{\sc Chaibub~Neto} {\em et~al.\/}(2008){\sc Chaibub~Neto}, {\sc
  Ferrara}, {\sc Attie} and {\sc Yandell}}]{Neto:2008}
{\sc Chaibub~Neto, E.}, {\sc C.~T. Ferrara}, {\sc A.~D. Attie}, and {\sc B.~S.
  Yandell}, 2008 Inferring causal phenotype networks from segregating
  populations.
\newblock Genetics {\bf 179}: 1089--100.

\bibitem[{{\sc Chaibub~Neto} {\em et~al.\/}(2010){\sc Chaibub~Neto}, {\sc
  Keller}, {\sc Attie} and {\sc Yandell}}]{Neto:2010}
{\sc Chaibub~Neto, E.}, {\sc M.~P. Keller}, {\sc A.~D. Attie}, and {\sc B.~S.
  Yandell}, 2010 Causal graphical models in systems genetics: a unified
  framework for joint inference of causal network and genetic architecture for
  correlated phenotypes.
\newblock Ann Appl Stat {\bf 4}: 320--339.

\bibitem[{{\sc Chen} {\em et~al.\/}(2007){\sc Chen}, {\sc Emmert-Streib} and
  {\sc Storey}}]{Chen:2007}
{\sc Chen, L.~S.}, {\sc F.~Emmert-Streib}, and {\sc J.~D. Storey}, 2007
  Harnessing naturally randomized transcription to infer regulatory
  relationships among genes.
\newblock Genome Biol {\bf 8}: R219.

\bibitem[{{\sc Cheung} and {\sc Spielman}(2009)}]{CheungSpielman:2009}
{\sc Cheung, V.~G.}, and {\sc R.~S. Spielman}, 2009 Genetics of human gene
  expression: mapping {DNA} variants that influence gene expression.
\newblock Nat Rev Genet {\bf 10}: 595--604.

\bibitem[{{\sc Chin} {\em et~al.\/}(2008){\sc Chin}, {\sc Chubukov}, {\sc
  Jolly}, {\sc DeRisi} and {\sc Li}}]{Chin:2008}
{\sc Chin, C.-S.}, {\sc V.~Chubukov}, {\sc E.~R. Jolly}, {\sc J.~DeRisi}, and
  {\sc H.~Li}, 2008 Dynamics and design principles of a basic regulatory
  architecture controlling metabolic pathways.
\newblock PLoS Biol {\bf 6}: e146.

\bibitem[{{\sc Chun} and {\sc Kele{\c{s}}}(2009)}]{ChunKeles:2009}
{\sc Chun, H.}, and {\sc S.~Kele{\c{s}}}, 2009 Expression quantitative trait
  loci mapping with multivariate sparse partial least squares regression.
\newblock Genetics {\bf 182}: 79--90.

\bibitem[{{\sc Costanzo} {\em et~al.\/}(2010){\sc Costanzo}, {\sc
  Baryshnikova}, {\sc Bellay}, {\sc Kim}, {\sc Spear} {\em
  et~al.\/}}]{CostanzoBoone:2010}
{\sc Costanzo, M.}, {\sc A.~Baryshnikova}, {\sc J.~Bellay}, {\sc Y.~Kim}, {\sc
  E.~D. Spear}, {\em et~al.\/}, 2010 The genetic landscape of a cell.
\newblock Science {\bf 327}: 425--431.

\bibitem[{{\sc Curtis} {\em et~al.\/}(2013){\sc Curtis}, {\sc Kim}, {\sc
  Woolford~Jr}, {\sc Xu} and {\sc Xing}}]{Curtis:2013}
{\sc Curtis, R.~E.}, {\sc S.~Kim}, {\sc J.~L. Woolford~Jr}, {\sc W.~Xu}, and
  {\sc E.~P. Xing}, 2013 Structured association analysis leads to insight into
  {{Saccharomyces cerevisiae}} gene regulation by finding multiple contributing
  e{QTL} hotspots associated with functional gene modules.
\newblock BMC Genomics {\bf 14}: 1--17.

\bibitem[{{\sc Didelez} and {\sc Edwards}(2004)}]{Didelez:2004}
{\sc Didelez, V.}, and {\sc D.~Edwards}, 2004 Collapsibility of graphical
  cg-regression models.
\newblock Scand J Stat {\bf 31}: 535--551.

\bibitem[{{\sc Edwards}(2000)}]{Edwards:2000}
{\sc Edwards, D.}, 2000 {\em Introduction to graphical modelling\/}.
\newblock Springer.

\bibitem[{{\sc Edwards} {\em et~al.\/}(2010){\sc Edwards}, {\sc de~Abreu} and
  {\sc Labouriau}}]{Edwards:2010}
{\sc Edwards, D.}, {\sc G.~C.~G. de~Abreu}, and {\sc R.~Labouriau}, 2010
  Selecting high-dimensional mixed graphical models using minimal aic or bic
  forests.
\newblock BMC Bioinformatics {\bf 11}: 18.

\bibitem[{{\sc Grone} {\em et~al.\/}(1984){\sc Grone}, {\sc Johnson}, {\sc
  S{\'a}} and {\sc Wolkowicz}}]{GroneWolkowicz:1984}
{\sc Grone, R.}, {\sc C.~Johnson}, {\sc E.~S{\'a}}, and {\sc H.~Wolkowicz},
  1984 Positive definite completions of partial {H}ermitian matrices.
\newblock Linear Algebra Appl {\bf 58}: 109--124.

\bibitem[{{\sc Hanson}(2014)}]{Hanson:2014}
{\sc Hanson, B.~A.}, 2014 {\em {H}ive{R}: 2{D} and 3{D} {H}ive plots for
  {R}\/}.
\newblock R/CRAN pkg. ver. 0.2-27.

\bibitem[{{\sc Hastie} {\em et~al.\/}(2009){\sc Hastie}, {\sc Tibshirani} and
  {\sc Friedman}}]{HastieFriedman:2009}
{\sc Hastie, T.}, {\sc R.~Tibshirani}, and {\sc J.~Friedman}, 2009 {\em The
  elements of statistical learning\/}.
\newblock Springer.

\bibitem[{{\sc Hughes} {\em et~al.\/}(2000){\sc Hughes}, {\sc Marton}, {\sc
  Jones}, {\sc Roberts}, {\sc Stoughton} {\em et~al.\/}}]{Hughes:2000}
{\sc Hughes, T.~R.}, {\sc M.~J. Marton}, {\sc A.~R. Jones}, {\sc C.~J.
  Roberts}, {\sc R.~Stoughton}, {\em et~al.\/}, 2000 Functional discovery via a
  compendium of expression profiles.
\newblock Cell {\bf 102}: 109--126.

\bibitem[{{\sc Jansen} and {\sc Nap}(2001)}]{JansenNap:2001}
{\sc Jansen, R.~C.}, and {\sc J.-P. Nap}, 2001 Genetical genomics: the added
  value from segregation.
\newblock Trends Genet {\bf 17}: 388--390.

\bibitem[{{\sc Kalisch} and {\sc B{\"u}hlmann}(2007)}]{KalischBuhlmann:2007}
{\sc Kalisch, M.}, and {\sc P.~B{\"u}hlmann}, 2007 Estimating high-dimensional
  directed acyclic graphs with the pc-algorithm.
\newblock J Mach Learn Res {\bf 8}: 613--636.

\bibitem[{{\sc Kang} {\em et~al.\/}(2010){\sc Kang}, {\sc Ye}, {\sc Shpitser}
  and {\sc Eskin}}]{Kang:2010}
{\sc Kang, E.~Y.}, {\sc C.~Ye}, {\sc I.~Shpitser}, and {\sc E.~Eskin}, 2010
  Detecting the presence and absence of causal relationships between expression
  of yeast genes with very few samples.
\newblock J Comput Biol {\bf 17}: 533--46.

\bibitem[{{\sc Kang} {\em et~al.\/}(2008){\sc Kang}, {\sc Ye} and {\sc
  Eskin}}]{Kang:2008}
{\sc Kang, H.~M.}, {\sc C.~Ye}, and {\sc E.~Eskin}, 2008 Accurate discovery of
  expression quantitative trait loci under confounding from spurious and
  genuine regulatory hotspots.
\newblock Genetics {\bf 180}: 1909--25.

\bibitem[{{\sc Kendziorski} {\em et~al.\/}(2006){\sc Kendziorski}, {\sc Chen},
  {\sc Yuan}, {\sc Lan} and {\sc Attie}}]{KendziorskiAttie:2006}
{\sc Kendziorski, C.}, {\sc M.~Chen}, {\sc M.~Yuan}, {\sc H.~Lan}, and {\sc
  A.~Attie}, 2006 Statistical methods for expression quantitative trait loci
  (e{QTL}) mapping.
\newblock Biometrics {\bf 62}: 19--27.

\bibitem[{{\sc Kim} and {\sc Xing}(2009)}]{Kim:2009}
{\sc Kim, S.}, and {\sc E.~P. Xing}, 2009 Statistical estimation of correlated
  genome associations to a quantitative trait network.
\newblock PLoS Genet {\bf 5}: e1000587.

\bibitem[{{\sc Krzywinski} {\em et~al.\/}(2012){\sc Krzywinski}, {\sc Birol},
  {\sc Jones} and {\sc Marra}}]{Krzywinski:2012}
{\sc Krzywinski, M.}, {\sc I.~Birol}, {\sc S.~J. Jones}, and {\sc M.~A. Marra},
  2012 Hive plots--rational approach to visualizing networks.
\newblock Brief Bioinform {\bf 13}: 627--644.

\bibitem[{{\sc Lauritzen}(1996)}]{Lauritzen:1996}
{\sc Lauritzen, S.}, 1996 {\em Graphical Models\/}.
\newblock Oxford University Press.

\bibitem[{{\sc Lauritzen} and {\sc Wermuth}(1989)}]{Lauritzen:1989}
{\sc Lauritzen, S.}, and {\sc N.~Wermuth}, 1989 Graphical models for
  associations between variables, some of which are qualitative and some
  quantitative.
\newblock Ann Stat {\bf 17}: 31--57.

\bibitem[{{\sc Lee} {\em et~al.\/}(2011){\sc Lee}, {\sc Wray}, {\sc Goddard}
  and {\sc Visscher}}]{LeeVisscher:2011}
{\sc Lee, S.~H.}, {\sc N.~R. Wray}, {\sc M.~E. Goddard}, and {\sc P.~M.
  Visscher}, 2011 Estimating missing heritability for disease from genome-wide
  association studies.
\newblock Am J Hum Genet {\bf 88}: 294--305.

\bibitem[{{\sc Leek} {\em et~al.\/}(2010){\sc Leek}, {\sc Scharpf}, {\sc
  Bravo}, {\sc Simcha}, {\sc Langmead} {\em et~al.\/}}]{LeekIrizarry:2010}
{\sc Leek, J.~T.}, {\sc R.~B. Scharpf}, {\sc H.~C. Bravo}, {\sc D.~Simcha},
  {\sc B.~Langmead}, {\em et~al.\/}, 2010 Tackling the widespread and critical
  impact of batch effects in high-throughput data.
\newblock Nat Rev Genet {\bf 11}: 733--739.

\bibitem[{{\sc Leek} and {\sc Storey}(2007)}]{LeekStorey:2007}
{\sc Leek, J.~T.}, and {\sc J.~D. Storey}, 2007 Capturing heterogeneity in gene
  expression studies by surrogate variable analysis.
\newblock PLoS Genet {\bf 3}: e161.

\bibitem[{{\sc Li} {\em et~al.\/}(2002){\sc Li}, {\sc Peterson}, {\sc Fang} and
  {\sc Stamatoyannopoulos}}]{LiStam:2002}
{\sc Li, Q.}, {\sc K.~R. Peterson}, {\sc X.~Fang}, and {\sc
  G.~Stamatoyannopoulos}, 2002 Locus control regions.
\newblock Blood {\bf 100}: 3077--3086.

\bibitem[{{\sc Listgarten} {\em et~al.\/}(2010){\sc Listgarten}, {\sc Kadie}
  and {\sc Heckerman}}]{Listgarten:2010}
{\sc Listgarten, J.}, {\sc C.~Kadie}, and {\sc D.~Heckerman}, 2010 {Correction
  for hidden confounders in the genetic analysis of gene expression}.
\newblock P Natl Acad Sci USA {\bf 107}: 16465--16470.

\bibitem[{{\sc Liu} {\em et~al.\/}(2008){\sc Liu}, {\sc de~la Fuente} and {\sc
  Hoeschele}}]{Liu:2008}
{\sc Liu, B.}, {\sc A.~de~la Fuente}, and {\sc I.~Hoeschele}, 2008 Gene network
  inference via structural equation modeling in genetical genomics experiments.
\newblock Genetics {\bf 178}: 1763--76.

\bibitem[{{\sc Michaelson} {\em et~al.\/}(2010){\sc Michaelson}, {\sc Alberts},
  {\sc Schughart} and {\sc Beyer}}]{MichaelsonBeyer:2010}
{\sc Michaelson, J.~J.}, {\sc R.~Alberts}, {\sc K.~Schughart}, and {\sc
  A.~Beyer}, 2010 Data-driven assessment of e{QTL} mapping methods.
\newblock BMC genomics {\bf 11}: 502.

\bibitem[{{\sc Montgomery} {\em et~al.\/}(2010){\sc Montgomery}, {\sc Sammeth},
  {\sc Gutierrez-Arcelus}, {\sc Lach}, {\sc Ingle} {\em
  et~al.\/}}]{MontgomeryDermitzakis:2010}
{\sc Montgomery, S.~B.}, {\sc M.~Sammeth}, {\sc M.~Gutierrez-Arcelus}, {\sc
  R.~P. Lach}, {\sc C.~Ingle}, {\em et~al.\/}, 2010 Transcriptome genetics
  using second generation sequencing in a caucasian population.
\newblock Nature {\bf 464}: 773--777.

\bibitem[{{\sc Parts} {\em et~al.\/}(2011){\sc Parts}, {\sc Stegle}, {\sc Winn}
  and {\sc Durbin}}]{Parts:2011}
{\sc Parts, L.}, {\sc O.~Stegle}, {\sc J.~Winn}, and {\sc R.~Durbin}, 2011
  Joint genetic analysis of gene expression data with inferred cellular
  phenotypes.
\newblock PLoS Genet {\bf 7}: e1001276.

\bibitem[{{\sc Petretto} {\em et~al.\/}(2006){\sc Petretto}, {\sc Mangion},
  {\sc Dickens}, {\sc Cook}, {\sc Kumaran} {\em
  et~al.\/}}]{PetrettoAitman:2006}
{\sc Petretto, E.}, {\sc J.~Mangion}, {\sc N.~J. Dickens}, {\sc S.~A. Cook},
  {\sc M.~K. Kumaran}, {\em et~al.\/}, 2006 Heritability and tissue specificity
  of expression quantitative trait loci.
\newblock PLoS Genet {\bf 2}: e172.

\bibitem[{{\sc Rao}(1973)}]{Rao:1973}
{\sc Rao, C.}, 1973 {\em {Linear Statistical Inference and Its
  Applications}\/}.
\newblock John Wiley \& Sons.

\bibitem[{{\sc Ritchie} {\em et~al.\/}(2007){\sc Ritchie}, {\sc Silver}, {\sc
  Oshlack}, {\sc Holmes}, {\sc Diyagama} {\em et~al.\/}}]{Ritchie:2007}
{\sc Ritchie, M.~E.}, {\sc J.~Silver}, {\sc A.~Oshlack}, {\sc M.~Holmes}, {\sc
  D.~Diyagama}, {\em et~al.\/}, 2007 A comparison of background correction
  methods for two-colour microarrays.
\newblock Bioinformatics {\bf 23}: 2700--2707.

\bibitem[{{\sc Rockman}(2008)}]{Rockman:2008}
{\sc Rockman, M.~V.}, 2008 Reverse engineering the genotype--phenotype map with
  natural genetic variation.
\newblock Nature {\bf 456}: 738--744.

\bibitem[{{\sc Rockman} and {\sc Kruglyak}(2006)}]{RockmanKruglyak:2006}
{\sc Rockman, M.~V.}, and {\sc L.~Kruglyak}, 2006 Genetics of global gene
  expression.
\newblock Nat Rev Genet {\bf 7}: 862--872.

\bibitem[{{\sc Roverato}(2002)}]{Roverato:2002}
{\sc Roverato, A.}, 2002 {Hyper inverse {W}ishart distribution for
  non-decomposable graphs and its application to {B}ayesian inference for
  {G}aussian graphical models}.
\newblock Scand J Stat {\bf 29}: 391--411.

\bibitem[{{\sc Schadt} {\em et~al.\/}(2003){\sc Schadt}, {\sc Monks}, {\sc
  Drake}, {\sc Lusis}, {\sc Che} {\em et~al.\/}}]{SchadtFriend:2003}
{\sc Schadt, E.~E.}, {\sc S.~A. Monks}, {\sc T.~A. Drake}, {\sc A.~J. Lusis},
  {\sc N.~Che}, {\em et~al.\/}, 2003 Genetics of gene expression surveyed in
  maize, mouse and man.
\newblock Nature {\bf 422}: 297--302.

\bibitem[{{\sc Seber}(2007)}]{Seber:2007}
{\sc Seber, G.}, 2007 {\em {A matrix handbook for statisticians}\/}.
\newblock Wiley-Interscience.

\bibitem[{{\sc Smyth} and {\sc Speed}(2003)}]{Smyth:2003}
{\sc Smyth, G.~K.}, and {\sc T.~Speed}, 2003 Normalization of c{DNA} microarray
  data.
\newblock Methods {\bf 31}: 265--273.

\bibitem[{{\sc Stegle} {\em et~al.\/}(2010){\sc Stegle}, {\sc Parts}, {\sc
  Durbin} and {\sc Winn}}]{StegleWinn:2010}
{\sc Stegle, O.}, {\sc L.~Parts}, {\sc R.~Durbin}, and {\sc J.~Winn}, 2010 A
  {B}ayesian framework to account for complex non-genetic factors in gene
  expression levels greatly increases power in e{QTL} studies.
\newblock PLoS Comp Biol {\bf 6}: e1000770.

\bibitem[{{\sc Teixeira} {\em et~al.\/}(2014){\sc Teixeira}, {\sc Monteiro},
  {\sc Guerreiro}, {\sc Gon{\c{c}}alves}, {\sc Mira} {\em
  et~al.\/}}]{Teixeira:2014}
{\sc Teixeira, M.~C.}, {\sc P.~T. Monteiro}, {\sc J.~F. Guerreiro}, {\sc J.~P.
  Gon{\c{c}}alves}, {\sc N.~P. Mira}, {\em et~al.\/}, 2014 The {YEASTRACT}
  database: an upgraded information system for the analysis of gene and genomic
  transcription regulation in {{Saccharomyces cerevisiae}}.
\newblock Nucleic Acids Res {\bf 42}: D161--D166.

\bibitem[{{\sc Tesson} and {\sc Jansen}(2009)}]{TessonJansen:2009}
{\sc Tesson, B.~M.}, and {\sc R.~C. Jansen}, 2009 e{QTL} analysis in mice and
  rats.
\newblock In {\em Cardiovascular Genomics\/}, volume 573 of {\em Methods in
  {M}olecular {B}iology\/}. Springer, 285--309.

\bibitem[{{\sc Westra} {\em et~al.\/}(2013){\sc Westra}, {\sc Peters}, {\sc
  Esko}, {\sc Yaghootkar}, {\sc Schurmann} {\em et~al.\/}}]{WestraFranke:2013}
{\sc Westra, H.-J.}, {\sc M.~J. Peters}, {\sc T.~Esko}, {\sc H.~Yaghootkar},
  {\sc C.~Schurmann}, {\em et~al.\/}, 2013 Systematic identification of trans
  e{QTL}s as putative drivers of known disease associations.
\newblock Nat Genet {\bf 45}: 1238--1243.

\bibitem[{{\sc Yvert} {\em et~al.\/}(2003){\sc Yvert}, {\sc Brem}, {\sc
  Whittle}, {\sc Akey}, {\sc Foss} {\em et~al.\/}}]{Yvert:2003}
{\sc Yvert, G.}, {\sc R.~B. Brem}, {\sc J.~Whittle}, {\sc J.~M. Akey}, {\sc
  E.~Foss}, {\em et~al.\/}, 2003 Trans-acting regulatory variation in
  {{Saccharomyces cerevisiae}} and the role of transcription factors.
\newblock Nat Genet {\bf 35}: 57--64.

\bibitem[{{\sc Zhu} {\em et~al.\/}(2004){\sc Zhu}, {\sc Lum}, {\sc Lamb}, {\sc
  GuhaThakurta}, {\sc Edwards} {\em et~al.\/}}]{Zhu:2004}
{\sc Zhu, J.}, {\sc P.~Y. Lum}, {\sc J.~Lamb}, {\sc D.~GuhaThakurta}, {\sc
  S.~W. Edwards}, {\em et~al.\/}, 2004 An integrative genomics approach to the
  reconstruction of gene networks in segregating populations.
\newblock Cytogenet Genome Res {\bf 105}: 363--74.

\bibitem[{{\sc Zhu} {\em et~al.\/}(2008){\sc Zhu}, {\sc Zhang}, {\sc Smith},
  {\sc Drees}, {\sc Brem} {\em et~al.\/}}]{Zhu:2008}
{\sc Zhu, J.}, {\sc B.~Zhang}, {\sc E.~N. Smith}, {\sc B.~Drees}, {\sc R.~B.
  Brem}, {\em et~al.\/}, 2008 Integrating large-scale functional genomic data
  to dissect the complexity of yeast regulatory networks.
\newblock Nat Genet {\bf 40}: 854--861.

\end{thebibliography}

\end{document}